\documentclass[11pt,a4paper]{article}

\usepackage{jheppub}
\usepackage[T1]{fontenc}
\usepackage{amssymb}
\usepackage{mathtools}
\usepackage{amsmath}
\usepackage{bm}
\usepackage{stackengine}

\usepackage{scalerel}
\usepackage{eufrak}
\usepackage{framed}

\usepackage{mathtools}

\usepackage{axodraw4j}
\usepackage{color}
\usepackage{pstricks}

\makeatletter
\newcommand{\doublewidetilde}[1]{{%
  \mathpalette\double@widetilde{#1}%
}}
\newcommand{\double@widetilde}[2]{%
  \sbox\z@{$\m@th#1\widetilde{#2}$}%
  \ht\z@=.9\ht\z@
  \widetilde{\box\z@}%
}
\makeatother

\usepackage{hyperref}
\usepackage{slashed}
\usepackage{epsfig}
\usepackage{amsmath,latexsym,amssymb}
\usepackage{graphicx}
\usepackage[latin1]{inputenc}
\usepackage{braket}
\usepackage{pdfsync}

\usepackage{framed}

\usepackage{mathtools}

\def\be{\begin{equation}}
\def\ee{\end{equation}}
\def\ba{\begin{eqnarray}}
\def\ea{\end{eqnarray}}

\newcommand{\bz}{\bar{z}}
\newcommand{\bw}{\bar{w}}
\newcommand{\bh}{\bar{h}}

\newcommand{\zbar}{\bar{z}}

\newcommand{\ha}{{1\over 2}}

\def\wbar{\bar w}

\def\d{\delta}
\def\D{\Delta}
\def\e{\epsilon}

\def\m{\mu}
\def\n{\nu}
\def\om{\omega}

\def\cM{{\cal M}}

\def\cO{{\cal O}}

\def\cA{{\cal A}}

\newcommand{\comment}[1]{}

\newcommand{\eea}{\end{eqnarray}}

\author{
Angelos Fotopoulos${}^{1,2}$, Stephan Stieberger${}^3$,
Tomasz R.\ Taylor${}^{1}$,\, Bin Zhu${}^1$\\[0.5cm]
 $^1${\it Department of Physics \\
  Northeastern University, Boston, MA 02115, USA}\\[0.2cm]
 $^2${\it Department of Sciences\\
Wentworth Institute of Technology, Boston, MA 02115, USA} \\[.2 cm] $^3${\it Max--Planck--Institut f\"ur Physik\\ Werner--Heisenberg--Institut, 80805 M\"unchen, Germany}
 }

\title{\boldmath Extended BMS Algebra of Celestial CFT \unboldmath}

\abstract{We elaborate on the proposal of flat holography in which four-dimensional physics is encoded in two-dimensional celestial conformal field theory (CCFT). The symmetry underlying CCFT is the extended BMS symmetry of (asymptotically) flat spacetime. We use soft and collinear theorems of Einstein-Yang-Mills theory to derive the OPEs of BMS field operators generating superrotations and supertranslations. The energy-momentum tensor, given by a shadow transform of a soft graviton operator, implements superrotations in the Virasoro subalgebra of $\mathfrak{bms_4}$. Supertranslations can be obtained from a single translation generator along the light-cone direction by commuting it with the energy-momentum tensor. This operator also originates from a soft graviton and generates a flow of conformal dimensions. All supertranslations can be assembled into a single primary conformal field operator on celestial sphere.
}


\bigskip

\begin{document}
\maketitle
\section{Introduction}
Bondi-van der Burg-Metzner-Sachs  (BMS) group is a symmetry of asymptotically flat four-dimensional spacetimes at null infinity \cite{Bondi:1962px,Sachs:1962wk}. It is a semi-direct product of the $SL(2,\mathbb{C})$ subgroup of global conformal transformations of the celestial sphere ${\cal C S}^2$ at null infinity (isomorphic to Lorentz transformations) times the abelian subgroup of so-called supertranslations. In 2009, Barnich and Troessaert \cite{Barnich:2009se} argued that $SL(2,\mathbb{C})$ should be extended to the group of all {\em local\/} conformal transformations (diffeomorphisms) of celestial sphere, a.k.a.\ superrotations. In 2013, Strominger showed that such extended BMS is also a symmetry  of the S-matrix describing elementary particles and gravitons \cite{Strominger:2013jfa}. This symmetry plays central role in Strominger's proposal of a flat spacetime hologram on ${\cal C S}^2$ \cite{Strominger:2017zoo}.\footnote{For a review of more recent developments, see  Ref.\cite{Pasterski:2019msg}. For an earlier proposal of flat holography, see Ref.\cite{deBoer:2003vf}. } In particular, superrotations set the stage for two-dimensional celestial conformal field theory (CCFT) on ${\cal C S}^2$ which, according to Strominger, should encode four-dimensional physics.

In CCFT, each particle is associated to a conformal field operator. The correlators of these operators are identified with  four-dimensional S-matrix elements transformed to  the basis of
conformal wave packets. They can be  obtained by applying Mellin transformations (with respect to the energies of external particles) to traditional, momentum space amplitudes
\cite{Pasterski:2016qvg,Pasterski:2017ylz,Schreiber:2017jsr,Stieberger:2018edy}.
Within this framework, the operator insertion points $z\in {\cal C S}^2$ are identified with the asymptotic directions of four-momenta while their dimensions $\Delta$ are identified with the dimensions of wave packets. The conformal wave packets associated to stable, helicity $\ell$ particles  have conformal weights $(h, \bar h)=(\frac{\Delta+\ell}{2},\frac{\Delta-\ell}{2})$ with Re$(\Delta)=1$ \cite{Pasterski:2017kqt}. Two of us have recently  shown that, given the proper definition of CCFT energy-momentum tensor, the operators representing gauge bosons and gravitons do indeed transform under diffeomorphisms of ${\cal C S}^2$ (superrotations) as primary conformal fields \cite{Fotopoulos:2019tpe}.

There are two special cases of conformal dimensions that lead to important insights into CCFT. For gauge bosons, it is the ``conformally soft'' limit of $\Delta=1$ \cite{Donnay:2018neh} in which  conformal wave packets describe pure gauge fields. They correspond to asymptotically ``large'' gauge configurations that have observable effects similar to  Goldstone modes. The operators emerging in the   $\Delta=1$ $(h=1,\,\bar h=0)$ limit of gauge boson operators are the holomorphic currents $J^a$ carrying gauge group charges. In Ref.\cite{FFT2019}, we derived OPEs of $J^a$ with other operators and showed that respective Ward identities agree with soft theorems. For gravitons, both $\Delta=0$ and $\Delta=1$ spin 2 wave packets represent (large) diffeomorphisms. Upon a shadow transformation, the operator associated to $\Delta=0\; (h=1,\bar h=-1)$ graviton (which is outside the Re$(\Delta)=1$ stability domain) changes its dimension to $\D=2\;  (h=2,\bar h=0)$ and becomes the CCFT energy-momentum tensor $T$ generating superrotations \cite{Kapec:2016jld,Cheung:2016iub}. The graviton associated to $\D=1\; (h=3/2,\bar h=-1/2)$ yields a primary field operator with a single-derivative descendant $P$ that generates supertranslations. Hence BMS symmetries are controlled by $\Delta=0$ and $\Delta=1$ soft limits. On the other hand, collinear limits are  useful for studying OPEs because identical momentum directions correspond to the operator insertion points coinciding on ${\cal C S}^2$. We will utilize them together with the soft limits in the following discussion of the extended  BMS symmetry algebra $\mathfrak{bms_4}$.

While BMS is a highly nontrivial symmetry of asymptotically flat spacetimes, its implementation in CCFT is rather straightforward. Virasoro subalgebra must emerge from the standard OPE of  $TT$ products. We also know what to expect from the OPE of the supertranslation operator $P$ with $T$ because $P$ is a well-defined descendant of a primary field. What is non-trivial in the context of CCFT is to show that such OPEs follow from the properties of Einstein-Yang-Mills (EYM) theory of gauge bosons coupled to gravitons. In this work, we extract $\mathfrak{bms_4}$ from the collinear and soft limits of EYM amplitudes.

The paper is organized as follows. In Section 2, we revisit OPEs of the operators associated to gauge bosons and gravitons previously discussed in Refs.\cite{FFT2019} and \cite{Pate:2019lpp}. We rewrite them in uniform normalization coventions. The relevant collinear limits are collected in Appendix A, where they are derived by using EYM Feynman rules. In Section 3, we derive the OPEs of superrotations generated by the energy-momentum tensor and of superstranslations generated by a descendant of a soft graviton operator, with the operators associated to gauge bosons and gravitons. Although the energy-momentum tensor is defined by a non-local shadow integral, its OPEs are always localized. In Section 4, we use soft theorems \cite{FFT2019,Pate:2019mfs,Nandan:2019jas,
Adamo:2019ipt,Puhm:2019zbl,Guevara:2019ypd} to derive the OPEs of BMS generators.\footnote{These OPEs have been discussed before in Ref.\cite{distler} by using different methods.}  In Section 5 we connect   OPEs with $\mathfrak{bms_4}$.

\label{intro}
\section{Preliminaries: OPEs of spin 1 and 2 operators}
The connection between light-like four-momenta $p^\mu$ of massless particles and points $z\in {\cal C S}^2$ relies on the following  parametrization:
\be\label{momparam}
p^\mu= \omega q^\mu, \qquad q^\mu={1\over 2} (1+|z|^2, z+\bz, -i(z-\bz), 1-|z|^2)\ ,
\ee
where $\omega$ is the light-cone energy and $q^\mu$ is a null vector -- the direction along which the massless state propagates,  determined by $z$.
The basis of wave functions required for transforming scattering amplitudes into CCFT correlators consist of conformal wave packets characterized by $z$, dimension $\D$ and helicity $\ell$.
The starting point for constructing such packets are Mellin transforms of spin 1 and 2 plane waves:
\be\label{confprimMellin1}
V^{\D , \ell}_{\mu} (X^\mu, z, \bz)\equiv \partial_J q_\mu \int _0^\infty d\omega  \ \omega^{\D-1} e^{\mp i \omega q \cdot  X -\e \omega} \qquad\quad~~~ (\ell=\pm 1),
\ee
\be\label{confprimMellin2}
H^{\D,\ell}_{\mu \n} (X^\mu, z, \bz)\equiv \partial_J q_\mu \partial_J q_\nu \int _0^\infty d\omega  \ \omega^{\D-1} e^{\mp i \omega q \cdot  X -\e \omega}\qquad(\ell=\pm 2) .
\ee
where $\partial_J=\partial_z$ for $\ell=+1,+2$ and $\partial_J=\partial_{\bz}$ for $\ell=-1,-2$.
The conformal (quasi-primary) wave functions can be written as
\be\label{confprimexpr3}
A^{\D,  \ell}_{\m }= g(\D){V^{\D ,\ell}_{\mu J} }+ \makebox{gauge}
\ ,\qquad
G^{\D,\ell}_{\m \n }=
f(\D)H^{\D,\ell}_{\mu \n} +\makebox{diff}
\ee
with the normalization constants
\be\label{gdef}g(\D)={\D -1 \over  \Gamma(\D+1) }\ ,\qquad f(\D)=
{1\over 2}{\D(\D-1) \over \Gamma(\D+2)} .\ee
The presence of these normalization factors makes it clear that, as mentioned in the Introduction, fields with spin 1 become pure gauge when $\D=1$ while fields with spin 2 become pure diffeomorphisms for $\D=0,1$.

The CCFT correlators are identified with the S-matrix elements transformed from the plane wave basis  into conformal basis (\ref{confprimexpr3}) by using properly normalized Mellin transformations \cite{Pasterski:2016qvg,Pasterski:2017ylz,Schreiber:2017jsr,Stieberger:2018edy}:
\ba\label{corrdef}
\Big\langle \prod_{n=1}^N\cO_{\D_n,\ell_n}(z_n,\zbar_n)\Big\rangle&=&\Big(  \prod_{n=1}^N c_n(\D_n) \int d \omega_n  \ \omega_n^{\D_n-1} \Big)  \d^{(4)}\big(\sum_{n=1}^N  \epsilon_n\om_n q_n\big)\nonumber\\[1mm] \label{cftcor}
 &&~~~~~\times
 \cM_{\ell_1\dots \ell_N}(\omega_n, z_n, \bz_n)
\ea
where $\cM_{\ell_1\dots \ell_N}$ are EYM Feynman's matrix elements with helicities $\ell_n$ and $c_n$ are the normalization constants
\be c_n(\D_n)=\left\{ ~{g(\D_n)~~\, \makebox{for}~~\ell_n=\pm 1,\atop f(\D_n) ~~\makebox{for}~~\ell_n=\pm 2 ,}\right. \label{cnns}\ee
see Eq.(\ref{gdef}). In Eq.(\ref{corrdef}) $\epsilon_n= +1$ or $-1$ depending whether the particles are incoming or outgoing, respectively. We skipped the gauge group indices and the corresponding group factors that can always be written in the basis of single or multiple trace Chan-Paton factors.

OPEs can be extracted from the correlators (\ref{corrdef}) by considering the limits of coinciding insertion points, which on the r.h.s.\ correspond to the collinear limits of scattering amplitudes.  Furthermore, in the cases of operators with $\D=1$ and 0, the zeros of normalization constants $c_n$ must be canceled  by  ``soft'' poles. Indeed, finite OPE coefficients appear from such  singular soft and collinear limits.

In Ref.\cite{FFT2019}, we derived the OPE coefficients of the products of spin 1 operators representing gauge bosons. They follow from the well-known collinear limits of  Yang-Mills amplitudes. {}For two gauge bosons labeled by gauge indices $a$ and $b$, with identical helicities,  we obtained
\be\label{opepp}
\cO^{a}_{\D_1, +}(z,\zbar)\,\cO^{b}_{\D_2,+}(w,\bar w)
=\frac{C_{(+,+)+}(\D_1,\D_2)}{z-w}\sum_c f^{abc}\cO^{c}_{(\D_1+\D_2-1),+}(w,\bar w)+\makebox{regular}\, ,
\ee
with
\be C_{(+,+)+}(\D_1,\D_2)=1-\frac{(\D_1-1)(\D_2-1)}{\D_1\D_2}\label{ope1}\ee
and a similar expression with  $(\bz-\bw)^{-1}$ pole  and the same OPE coefficient for two $-1$ helicities.

Two gauge bosons of opposite helicities can fuse into a single operator in the same way as in the case of identical helicities. The form of the corresponding OPE terms \cite{FFT2019} follows from the collinear limit of gauge interactions. In the case of opposite helicities, however, gravitational interactions allow the fusion of two gauge bosons into a graviton operator \cite{Pate:2019lpp}.
As explained in Appendix A, EYM amplitudes involving gravitational couplings do not blow up in the collinear limit but contain another type of singularity. It is due to a phase ambiguity reflecting an azimuthal asymmetry. In Appendix A, we show that when $z_1\to z_2$,
\begin{eqnarray}
{\cal M}(1^+,2^-,3,\dots, N) &=& \nonumber \frac{1}{\bar{z}_{12}}\frac{\omega_1}{\omega_2\omega_P}
{\cal M}(P^+,3,\dots , N)\,+\,\frac{1}{z_{12}}\frac{\omega_2}{\omega_1\omega_P}
{\cal M}(P^-,3,\dots ,N)\\[2mm]\nonumber
&&\,-\,\frac{z_{12}}{\bar{z}_{12}}\frac{\omega_1^2}{\omega_P^2}
{\cal M}(P^{++},3,\dots, N)\,-\,\frac{\bar{z}_{12}}{z_{12}}\frac{\omega_2^2}{\omega_P^2}
{\cal M}(P^{--},3,\dots ,N)\\[2.3mm] &&\,+\,\makebox{regular}\ .\label{eymcol}
\end{eqnarray}
Here, $P$ is the combined momentum of the collinear pair:
 \be \label{psum}P=p_1+p_2\equiv \omega_Pq_P\ ,\ee
 with
 \be\label{psum1}\om_P=\om_1+\om_2\, \qquad q_P=q_1=q_2~~(z_P=z_1=z_2,~\bz_P=\bz_1=\bz_2)\ ,\ee
and
\be z_{ij}\equiv z_i-z_j~,\qquad \bz_{ij}=\bz_i-\bz_j\ .\ee
 In Eq.(\ref{eymcol}), the last two terms, which originate from gravitational couplings,\footnote{In our units, the gravitational and gauge coupling constants $\kappa=2$ and $g_{\rm YM}=1$, respectively.}  contain the phase factors $z_{12}/\bz_{12}$ and $\bz_{12}/z_{12}$ depending on the direction from which $z_1$ approaches $z_2$ or equivalently, on the azimuthal angle of the plane spanned by the collinear pair about the axis of the combined momentum $P$. The ``regular'' part does not contain neither  $z_{12}$ nor $\bz_{12}$ in denominators and is well-defined, finite in the  $z_1= z_2$ limit. By following the same steps as in Ref.\cite{FFT2019}, we obtain
\ba
\cO^{a}_{\D_1 ,-}(z,\zbar)&&\!\!\!\!\cO^{b}_{\D_2,+}(w,\bar w)=\nonumber\\[2.5mm]
&&~~\frac{C_{(-,+)-}(\D_1,\D_2)}{z-w}\sum_c f^{abc}\cO^{c}_{(\D_1+\D_2-1),-}(w,\bar w)~~~~\label{opemp}\\[1.1mm]
&&+\,\frac{C_{(-+)+}(\D_1,\D_2)}{\zbar-\bar w}\sum_c f^{abc}\cO^{c}_{(\D_1+\D_2-1),+}(w,\bar w)\nonumber\\[.8mm]&& +\;C_{(-+)--}(\D_1,\D_2)\frac{\bar z-\bar w}{z-w}\,\delta^{ab}\,{\cal O}_{(\D_1+\D_2),-2}(w,\bar w)\nonumber\\[2mm]
&&+\; C_{(-+)++}(\D_1,\D_2)\frac{z-w}{\bar z-\bar w}\,\delta^{ab}\,{\cal O}_{(\D_1+\D_2),+2}(w,\bar w)\, +\,\makebox{regular}\, ,\nonumber
\ea
with
\ba C_{(-,+)-}(\D_1,\D_2)&=&\frac{\D_1-1}{\D_2(\D_1
+\D_2-2)}\label{ope3}\ ,\\[1mm]
C_{(-,+)+}(\D_1,\D_2)&=&\frac{\D_2-1}{\D_1(\D_1
+\D_2-2)}\label{ope4}\ ,
\\[2mm]
C_{(-+)--}(\D_1,\D_2)&=& - \frac{2(\D_1-1)(\D_1+1)(\D_2-1)}{\D_2(\D_1+\D_2)(\D_1+\D_2-1)}          \ , \\[2mm]
C_{(-+)++}(\D_1,\D_2) &=& - \frac{2(\D_2-1)(\D_2+1)(\D_1-1)}{\D_1(\D_1+\D_2)(\D_1+\D_2-1)}          \ .
\ea
The above result agrees with Refs.\cite{FFT2019,Pate:2019lpp}.

The OPEs involving gravitons can be extracted in a similar way. For two gravitons with identical helicities, we show in Appendix A that
\begin{equation}
\cM(1^{++},2^{++}, 3,\dots,N) = \frac{\omega_P^2}{\omega_{1}\omega_2}\frac{\bz_{12}}{z_{12}}
\cM(P^{++},3,\dots,N)+\makebox{regular}\, .\label{twogr}
\end{equation}
As a result,
 \begin{equation}\label{oprodh}
\cO_{\D_{1},+2}(z,\bar{z}) \cO_{\D_{2},+2}(w,\bw)  = C_{(+2,+2)+2}\frac{\bar{z}-\bar{w}}{z-w}\cO_{(\D_{1}+\D_{2}),+2}(w,\bw)
+\makebox{regular},
\end{equation}
with
\begin{equation}\label{eq:C+++}
C_{(+2,+2)+2} ={(\D_1+\D_{2}-2)(\D_1+\D_{2}+1)\over 2(\D_1+1)(\D_{2}+1)}.
\end{equation}
and a similar expression   with the conjugated phase factor and the same OPE coefficient for two $-2$ helicities.

The collinear limit of two gravitons with opposite helicities is also derived in Appendix A. It reads:
\ba
\cM(1^{--},2^{++}, 3,\dots,N) = && \frac{\omega_{1}^3}{\omega_{P}^2\omega_{2}}\frac{\bz_{12}}{z_{12}}
\cM(P^{--}, 3,\dots,N) \nonumber\\ &&+~
\frac{\omega_{2}^3}{\omega_{P}^2\omega_1}\frac{z_{12}}{\bz_{12}}
\cM(P^{++}, 3,\dots,N)+\makebox{regular}. \label{clgrav2}
\ea
The corresponding OPE is
\ba
\cO_{\D_{1},-2}(z,\bar{z}) \cO_{\D_{2},+2}(w,\bar{w})   = && C_{(-2,+2)-2}\frac{\bar{z}-\bar{w}}{z-w}\cO_{(\D_{1}+\D_2 ),-2}(w,\bar{w})\label{oprodg}\\
&& +~ C_{(-2,+2)+2}\frac{z-w}{\bar{z}-\bar{w}}\cO_{(\D_{1}+\D_2 ),+2}(w,\bw)+\makebox{regular},\nonumber
\ea
with
\ba\label{eq:Cmixed}
C_{(-2,+2)-2}&=&\frac{1}{2}{\D_{1}(\D_{1}-1)(\D_{1}+2) \over (\D_2+1)(\D_1+\D_{2}-1)(\D_1+\D_{2})} \ ,\\
C_{(-2,+2)+2}&=& \frac{1}{2}{\D_{2}(\D_{2}-1)(\D_{2}+2) \over (\D_{1}+1)(\D_1+\D_{2}-1)(\D_1+\D_2)}  \ .\label{eq:Cmixed2}
\ea

We close this section by listing the OPEs of gauge bosons with gravitons. They follow from the collinear limits of EYM amplitudes discussed in Appendix A:
\ba
\mathcal{M}(1^{++},2^{+},3,\dots,N) &=& \frac{\omega_P}{\omega_1}\frac{\bar{z}_{12}}{z_{12}}
\mathcal{M}(P^{+},3,\dots,N) \label{ggg1}+\makebox{regular},\label{spp11}\\
\mathcal{M}(1^{--},2^{+},3,\dots,N) &=& \frac{\omega_{2}^2}{\omega_{P}\omega_1}\frac{z_{12}}{\bz_{12}}
\mathcal{M}(P^+,3,\dots,N)\label{ggg2}+\makebox{regular},\label{spp12}
\ea
which lead to
\ba
\cO_{\D_{1},+}(z,\bar{z}) \cO_{\D_{2},+2}(w,\bw)  &=& C_{(+,+2)+}\frac{\bar{z}-\bar{w}}{z-w}\cO_{(\D_{1}+\D_{2}),+}
(w,\bar{w})+\makebox{regular},\label{ggg3}\\
\cO_{\D_{1},+}(z,\bar{z}) \cO_{\D_{2},-2}(w,\bw)  &=& C_{(+,-2),+}\frac{z-w}{\bz-\bar{w}}\cO_{(\D_{1}+\D_{2}),+}
(w,\bar{w})+\makebox{regular},\label{ggg4}
\ea
with
\ba\label{eq:CEY}
C_{(+,+2)+}&=&\frac{1}{2} \frac{(\Delta_{1}-1)(\Delta_{1} +\Delta_2)}{\Delta_{1}(\Delta_2 +1)}\label{ggg5}\ ,\\
C_{(+,-2)+}&=& \frac{1}{2}\frac{(\Delta_{1}-1)(\Delta_{1}+1)}{(\Delta_2 +1)(\Delta_{1}+\Delta_2-1)} \label{ggg6}\ .
\ea
The OPEs of operators with $+$ and $-$ interchanged have the phase factors $({z-w})/(\bz-\bar{w})$ conjugated and the same coefficients.

\section{OPEs of superrotations and supertranslations with spin 1 and spin
 2\\ operators}
In this section, we discuss the OPEs of  superrotations generated by the energy momentum tensor $T$ and of supertranslations generated by the operator $P$, with the operators associated to gauge bosons and gravitons. We know what to expect from the OPE of $T$ with primary fields. In Ref.\cite{Fotopoulos:2019tpe}, we showed that the operators $\cO^a_{\Delta,\pm}$ representing gauge bosons are indeed such primary fields. To that end, we used the energy-momentum tensor defined in \cite{Kapec:2016jld,Cheung:2016iub} as the shadow \cite{Osborn:2012vt}  of dimension $\Delta=0$ graviton operator $\cO_{0,-2}$:
\be\label{tdef}
T(z)\equiv\widetilde{\cal O}_{\D\to 0,-2}(z,\bz)={3!\over 2\pi} \int d^2 z'{1\over (z'-z)^{4}} \, {\cal O}_{0,-2}(z',\bz')
\ee
and took the collinear limit of the shadow  with gauge bosons as in Eq.(\ref{ggg2}). The limit of $\Delta\to 0$ was taken at the end. Actually, as we will see in this section, there is a fast way of deriving and generalizing the results of Ref.\cite{Fotopoulos:2019tpe}. We will first take the $\Delta\to 0$ limit and {\em then\/} take the shadow transform in order to extract the OPEs. We will proceed in the same way in the case of $P$:
\be\label{pdef} P(z,\bar z)\equiv\partial_{\bz} {\cal O}_{\D\to 1,+2}\ ,\ee
by taking the $\D\to 1$ limit first.
\subsection{Superrotations}
We start by inserting $T(z)$ into the correlator (\ref{corrdef}):
\ba\label{tcorr}
\Big\langle T(z) \prod_{n=1}^N\cO_{\D_n,\ell_n}(z_n,\zbar_n)\Big\rangle &=&\lim_{\Delta_0\to 0}\Big(  \prod_{n=0}^N c_n(\D_n)\Big){3!\over 2\pi} \int d^2 z_0{1\over (z_0-z)^{4}}
 \nonumber\\[1mm]
 &&~~~~\times\,
 \cA_{\ell_0=-2,\ell_1\dots \ell_N}(\D_n,z_n,\bz_n)
\ea
where we used the definition (\ref{tdef}) and
\ba\label{tmell}\cA_{\ell_0=-2,\ell_1\dots \ell_N}(\D_n,z_n,\bz_n)&\equiv&\cA_{-2}(\D_0,\D_1,\dots,\D_N)\\
=\Big( \prod_{n=0}^N \int d \omega_n  &&\!\!\!\!\!\!\!\! \omega_n^{\D_n-1}   \Big)\, \d^{(4)}\big(\sum_{n=0}^N  \epsilon_n\om_n q_n\big)\,
 \cM_{\ell_0=-2,\ell_1\dots \ell_N}(\omega_n, z_n, \bz_n)\ ,\nonumber
\ea which
is the Mellin-transformed amplitude with a ``soft'' graviton. Its $\D_0\to 0$ limit has been studied in Refs.\cite{Adamo:2019ipt,Puhm:2019zbl,Guevara:2019ypd}, with the following result:
\ba
\cA_{-2}&&\!\!\!\! (\D_0,\D_1,\dots,\D_N)\label{mason1} \\
&&\longrightarrow~\frac{1}{\Delta_0} \sum_{i=1}^N \frac{(z_0-z_{i})}{(\bz_0-\bz_{i})} \frac{(\bar\xi-\bz_i)}{(\bar\xi-\bz_0)}\Big[(z_0-z_i)\partial_{z_i} -2h_i\Big]\cA(\Delta_1,\dots,\Delta_N)\nonumber
\ea
where $\cA$ on the r.h.s.\ is the Mellin transform of the amplitude without the graviton. Here, $h_i$ is the chiral weight of the $i$th operator and $\xi$ is an arbitrary reference point on ${\cal C S}^2$ reflecting the gauge choice for the graviton. In this case, it is convenient to set $\xi\to\infty$. The shadow transform can be performed by writing
\be
{1\over (z_0-z)^{4}}=-\frac{1}{3!}\partial_{z_0}^3\Big({1\over z_0-z}\Big)\ee
and integrating by parts three times.
After repeatedly using the identity
\be \partial_{z_0}\Big({1\over \bz_0-\bz_i} \Big)=2\pi\delta^{(2)}(z_0-z_i)\ ,\ee
we obtain
\ba\label{tfcorr}
\Big\langle T(z)&& \!\!\!\!\prod_{n=1}^N\cO_{\D_n,\ell_n}(z_n,\zbar_n)\Big\rangle = \lim_{\Delta_0\to 0}\Big(  {c_0(\D_0)\over\D_0}\Big)
\\[1mm]
 &&~~~~\times\,\sum_{i=1}^N\Big[-\frac{2{h}_i}{(z-z_i)^2} - \frac{2}{z-z_i}{\partial}_{z_i}\Big]
 \Big\langle \prod_{n=1}^N\cO_{\D_n,\ell_n}(z_n,\zbar_n)\Big\rangle . \nonumber
\ea
Note that
\be\label{clim}
\lim_{\Delta_0\to 0}\Big(  {c_0(\D_0)\over\D_0}\Big)=\lim_{\Delta_0\to 0}\Big(  {f(\D_0)\over\D_0}\Big)=-\frac{1}{2}\ ,\ee
see Eq.(\ref{gdef}). As a result, we obtain
\begin{equation}\label{toope}
T(z)\cO_{\Delta,\ell}(w,\bw) = \frac{h}{(z-w)^2}\cO_{\Delta,\ell} (w,\bw) +\frac{1}{z-w}\partial_{w}\cO_{\Delta,\ell}(w,\bw) +\makebox{regular}\, .
\end{equation}
Following the same route for the antiholomorphic $\bar T(\bz)$, one obtains
\begin{equation}\label{btoope}
\bar T(\bz)\cO_{\Delta,\ell}(w,\bw) = \frac{\bh}{(\bz-\bw)^2}\cO_{\Delta,\ell} (w,\bw) +\frac{1}{\bz-\bw}\partial_{\bw}\cO_{\Delta,\ell}(w,\bw) +\makebox{regular}\, .
\end{equation}
The OPEs (\ref{toope}) and (\ref{btoope}), which are valid for both spin 1 and spin 2 particles, prove that the respective CCFT operators are primary fields.
\subsection{Supertranslations}

We start by inserting $P(z_0)$ into the correlator (\ref{corrdef}):
\be\label{pcorr}
\Big\langle P(z_0) \prod_{n=1}^N\cO_{\D_n,\ell_n}(z_n,\zbar_n)\Big\rangle =\lim_{\Delta_0\to 1}\Big(  \prod_{n=0}^N c_n(\D_n)\Big)\,\partial_{\bz_0}
 \cA_{\ell=+2,\ell_1\dots \ell_N}(\D_n,z_n,\bz_n)
\ee
where we used the definition (\ref{pdef}).
In this case, the relevant $\D_0\to 1$ soft limit is \cite{Adamo:2019ipt,Puhm:2019zbl,Guevara:2019ypd}:
\ba
\cA_{+2}&&\!\!\!\! (\D_0,\D_1,\dots,\D_N)\label{mason2} \\
&&\longrightarrow~\frac{1}{\Delta_0-1} \sum_{i=1}^N \frac{(\bz_0-\bz_{i})}{(z_0-z_{i})} \frac{(\xi-z_i)^2}{(\xi-z_0)^2}\cA(\Delta_1,\dots,\D_i+1,\dots,\Delta_N)\nonumber
\ea
Here again, as in Eq.(\ref{mason1}), we can set $\xi\to\infty$. {}From Eqs.(\ref{pcorr}) and (\ref{mason2}) it follows that
\ba\label{pfcorr}
\Big\langle P(z_0)&& \!\!\!\! \prod_{n=1}^N\cO_{\D_n,\ell_n}(z_n,\zbar_n)\Big\rangle =\lim_{\Delta_0\to 1}\Big(  { c_0(\D_0)\over\D_0-1}\Big)
~\sum_{i=1}^N\frac{c_i(\D_i)}{c_i(\D_i+1)}\frac{1}{z_0-z_i}\\[1mm]
 &&~\times
 \Big\langle \,\Big[ \prod_{m=1}^{i-1}\cO_{\D_m,\ell_m}(z_m,\zbar_m)\,\Big]\cO_{\D_i+1,\ell_i}(z_i,\zbar_i)
 \Big[\!\prod_{n=i+1}^N
 \cO_{\D_n,\ell_n}
 (z_n,\zbar_n)\,\Big]\,\Big\rangle .\nonumber
\ea
Note that
\be\lim_{\Delta_0\to 1}\Big(  { c_0(\D_0)\over\D_0-1}\Big)=\lim_{\Delta_0\to 1}\Big(  { f(\D_0)\over\D_0-1}\Big)=\frac{1}{4}\ee
and
\be \frac{c_i(\D_i)}{c_i(\D_{i}+1)}=\left\{ ~{\frac{g(\D_i)}{g(\D_{i}+1)}=\frac{(\D_i-1)(\D_i+1)}{\D_i}~~\, \makebox{for}~~\ell_i=\pm 1,\atop \frac{f(\D_i)}{f(\D_{i}+1)}=\frac{(\D_i-1)(\D_i+2)}{\D_i+1} ~~\makebox{for}~~\ell_i=\pm 2 ,}\right. \label{cnnis}\ee
As a result, we obtain
\ba\label{poope1}
P(z)\cO_{\Delta,\ell}(w,\bw)&=&\frac{(\D-1)(\D+1)}{4\D} \frac{1}{z-w}\cO_{\Delta+1,\ell}(w,\bw) +\makebox{regular}~~~(\ell=\pm 1),\\ \label{poope2}
P(z)\cO_{\Delta,\ell}(w,\bw)&=&\frac{(\D-1)(\D+2)}{4(\D+1)} \frac{1}{z-w}\cO_{\Delta+1,\ell}(w,\bw) +\makebox{regular}~~~(\ell=\pm 2),
\ea
and similar expressions for $\bar P(\bz)$, with complex conjugate poles.\footnote{The same result can be obtained from Eqs.(\ref{oprodh}), (\ref{oprodg}) and (\ref{ggg3}), derived in the previous section by considering the collinear instead of soft limits,  by taking the limit of $\D_2\to 1$.}
Note the presence of $(\D-1)$ factors in the above OPE coefficients. They imply that the products $P(z)J^a(w)$, $P(z)\bar J^a(\bw)$, $P(z)P(w)$ and  $P(z)\bar P(\bw)$ are regular.

$P(z)$ is not a holomorphic current: it is a dimension 2 descendant of a primary field, with $SL(2,\mathbb{C})$ weights (3/2,1/2). Nevertheless, the r.h.s.\ of Eq.(\ref{pfcorr}) implies the same Ward identity
as  translational symmetry of scattering amplitudes discussed in Refs.\cite{Stieberger:2018onx} and \cite{Law:2019glh}.
In fact, $P$ contains the $P_0+P_3$ component of the momentum operator which generates translations along the light-cone directions. The remaining components can be obtained by superrotations of $P$. This will be discussed in Section 5.
\section{OPEs of BMS generators}
\subsection{$TT$}
The product of energy-momentum tensors involves two shadow transforms. Let us define
\be
\doublewidetilde{\!\!\cA}(w,z,z_2,\dots)=\int d^2 z_0{1\over (z_0-w)^{4}} \int d^2 z_1{1\over (z_1-z)^{4}}\cA_{\ell_0=-2,\ell_1=-2,\ell_2\dots \ell_N}\ .\ee
Then
\be
\Big\langle T(w)T(z) \prod_{n=2}^N\cO_{\D_n,\ell_n}(z_n,\zbar_n)\Big\rangle =\bigg({3!\over 2\pi}\bigg)^2\lim_{\Delta_1\to 0}\lim_{\Delta_0\to 0}\bigg(\prod_{n=0}^N c_n(\D_n)\bigg)\;\,\doublewidetilde{\!\!\cA}(w,z,z_2,\dots)\ .\label{ttcorr}\ee
We take the limit $\D_0\to 0$ first, as in Eq.(\ref{mason1}):
\be
\cA_{\ell_0=-2,\ell_1=-2,\ell_1\dots \ell_N}\label{mason6}\longrightarrow~\frac{1}{\Delta_0} \sum_{i=1}^N \frac{(z_0-z_{i})}{(\bz_0-\bz_{i})} \frac{(\bar \xi-\bz_i)}{(\bar \xi-\bz_0)}\Big[(z_0-z_i)\partial_{z_i} -2h_i\Big]
\cA_{\ell_1=-2,\ell_2\dots \ell_N}\ ,
\ee
which leads to
\ba\label{step11} \doublewidetilde{\!\!\cA}(w,z,z_2,\dots)&\sim&
\int {d^2 z_0\over (z_0-w)^{4}}
  \int {d^2 z_1\over (z_1-z)^{4}}
 \frac{(z_0-z_{1})}{(\bz_0-\bz_{1})} \frac{(\bar \xi-\bz_1)}{(\bar \xi-\bz_0)}\Big[(z_0-z_1)\partial_{z_1} -2h_1^-\Big]
\cA_{\ell_1=-2,\ell_2\dots \ell_N}\nonumber\\[3mm]
&&\!\!\!\!\!\!\!\!\!\!\!+
 \Bigg( \int {d^2 z_0\over (z_0-w)^{4}}\sum_{i=2}^N
 \frac{(z_0-z_{i})}{(\bz_0-\bz_{i})} \frac{(\bar \xi-\bz_i)}{(\bar \xi-\bz_0)}\Big[(z_0-z_i)\partial_{z_i} -2h_i\Big]\Bigg)
\widetilde{\cA}(z,z_2,\dots)\ ,\label{step2}
  \ea
where $h_1^-=-1$ because $\ell_1=-2, \,\D_1=0$ and
\be \widetilde{\cA}(z,z_2,\dots)=\int {d^2 z_1\over (z_1-z)^{4}}\cA_{\ell_1=-2,\ell_2\dots \ell_N}\ .\ee
In Eq.(\ref{step11}), the integrals over $z_0$ can be performed by using the conformal integrals of Ref.\cite{Osborn1108}, see Appendix B. With the reference point $\xi\to\infty$, the result is
\ba
\lim_{\D_0\rightarrow 0}\frac{3!}{2\pi}c_0(\D_0)\,\,
\doublewidetilde{\!\!\cA}(w,z,z_2,\dots)\nonumber
&=&\int \frac{d^2 z_1}{(z-z_1)^4}\left( \frac{h^-_1}{(w-z_1)^2}+\frac{1}{w-z_1}\partial_{z_1}\right)\cA_{\ell_1=-2,\ell_2\dots \ell_N}\\
&&\!\!\!+\sum_{i =2}^N\left( \frac{h_i}{(w-z_i)^2}+\frac{1}{w-z_i}\partial_{z_i}\right)\widetilde{\cA}
(z,z_2,\dots)\ ,\label{ttfcorr}
\ea
The same result can be obtained by a direct application of Ward identity (\ref{tfcorr}).

It is clear from Eq.(\ref{ttfcorr}) that the second term (involving the sum over $i\ge 2$) is non-singular in the limit of $w\rightarrow z$, therefore only the first term needs to be included in the  derivation of OPE:
\ba\label{eq:TTstep1}
\langle T(w) T(z)&&\!\!\!\!\prod_{n=2}^N\mathcal{O}(z_n,\bar{z}_n)\rangle =\\&&
=\lim_{\Delta_1\rightarrow0}\frac{3!}{2\pi} \int \frac{d^2 z_1}{(z-z_1)^4}\left( \frac{1}{w-z_1}\partial_{z_1}-\frac{1}{(w-z_1)^2}\right)\nonumber
{\cal G}^-(z_1,\dots)\ , \ea
where \be{\cal G}^-(z_1,\dots)=\langle
\cO_{\Delta_1,-2}(z_1,\bz_1)\prod_{n=2}^N\mathcal{O}(z_n,\bar{z}_n)\rangle\ .\nonumber
\ee
In order to simplify notation, we introduce the following variables:
\begin{equation}
Z= z-z_1, \qquad W = w-z_1\ . \qquad 
\end{equation}
We anticipate a single pole term of the following form:
\ba\nonumber
\frac{1}{w-z}\langle \partial T &&\!\!\!\! \prod_{n=2}^N\mathcal{O}(z_n,\bar{z}_n)\rangle = \lim_{\Delta_1\rightarrow0}\frac{3!}{2\pi}
\frac{1}{w-z}\partial_z \int \frac{d^2z_1}{Z^4}{\cal G}^-(z_1,\dots)\\ \nonumber
&& =\lim_{\Delta_1\rightarrow0}\frac{3!}{2\pi}\left(-\frac{1}{w-z}\right)\int d^2z_1 \left( \partial_{z_1} \frac{1}{Z^4}\right){\cal G}^-(z_1,\dots)\\
&& =\lim_{\Delta_1\rightarrow0}\frac{3!}{2\pi}\frac{1}{w-z}\int \frac{d^2 z_1}{Z^4}\partial_{z_1}{\cal G}^-(z_1,\dots)\ ,
\ea
where in the last step we used integration by parts. Indeed, we can rewrite the r.h.s.\ of (\ref{eq:TTstep1}) as
\ba
\lim_{\Delta_1\rightarrow0}&&\frac{3!}{2\pi}\int d^2 z_1 \left( -\frac{1}{Z^4 W^2} +\frac{1}{Z^4W}\partial_{z_1}\right){\cal G}^-(z_1,\dots)= \nonumber\\
&&=\lim_{\Delta_1\rightarrow0}\frac{3!}{2\pi}\left[ -\frac{1}{Z^4W^2}+\left( \frac{1}{Z^4}-\frac{1}{Z^3W}\right)\frac{1}{w-z}\partial_{z_1}\right]{\cal G}^-(z_1,\dots)=\label{step21}\\
&&=-\lim_{\Delta_1\rightarrow0}\frac{3!}{2\pi}\left( \frac{1}{Z^4W^2}+\frac{1}{w-z}\frac{1}{Z^3W}\partial_{z_1}\right){\cal G}^-(z_1,\dots)+\frac{1}{w-z}\langle \partial T(z) \cdots\rangle\ .\nonumber
\ea
In the last line, the first term can be written as
\ba
\lim_{\Delta_1\rightarrow0}&&\frac{3!}{2\pi}\left[\frac{2}{(w-z)^2}\int \frac{d^2 z_1}{Z^4}{\cal G}^-(z_1,\dots) -\frac{2}{w-z}\int d^2 z_1 \left(\frac{1}{w-z}\frac{1}{Z^3 W}-\frac{1}{Z^3W^2}\right){\cal G}^-(z_1,\dots)\right]\nonumber\\
=&&\frac{2}{(w-z)^2}\langle T(z)\cdots\rangle-\lim_{\Delta_1\rightarrow0}\frac{3!}{2\pi}\frac{2}{(w-z)^2}\int d^2 z_1\frac{1}{Z^2W^2}{\cal G}^-(z_1,\dots)\ . \label{step22}
\ea
After combining Eqs.(\ref{step21}) and (\ref{step22}), we obtain:
\ba
\langle T(w) T(z)\prod_{n=2}^N\mathcal{O}(z_n,\bar{z}_n)\rangle &=&
\frac{2}{(w-z)^2}\langle \nonumber T(z)\prod_{n=2}^N\mathcal{O}(z_n,\bar{z}_n)\rangle+\frac{1}{w-z}\langle \partial T(z) \prod_{n=2}^N\mathcal{O}(z_n,\bar{z}_n)\rangle\\ &-&\lim_{\Delta_1\rightarrow0}\frac{3!}{2\pi}\frac{2}{(w-z)^2}\int d^2 z_1\frac{1}{Z^2W^2}{\cal G}^-(z_1,\dots)\ .\label{step31}
\ea
The final step is to show that the last term vanishes due to global conformal invariance. To that end, we take the soft limit $\Delta_1\to 0$:
\begin{equation}{\cal G}^-(z_1,\dots)\to\frac{c(\Delta_1)}{\Delta_1} \sum_{i=2}^N\frac{z_1-z_i}{\bar{z}_1-\bar{z}_i}\frac{\bar{\sigma}-
\bar{z}_i}{\bar{\sigma}-\bar{z}_1}[(z_1-z_i)\partial_i -2h_i]\langle\prod_{n=2}^N\mathcal{O}(z_n,\bar{z}_n)\rangle\ .
\end{equation}
After choosing the reference point $\sigma=z$, the  derivative terms become
\ba
\int \frac{d^2 z_1}{Z^2W^2}&&\!\!\!\!\!\!\frac{(z_1-z_i)^2}{(\bar{z}_1-\bar{z}_i)
(\bar{z}_1-\bar{z})}(\bar{z}_i-\bar{z})\partial_i\langle\prod_{n=2}^N
\mathcal{O}(z_n,\bar{z}_n)\rangle=\\
&=&\Gamma(0)\!\left(-\frac{1}{(z-w)^3}(z_i-w)^2\partial_i+\frac{1}{(z-w)^2}
(z_i-w)\partial_i \right)\langle\prod_{n=2}^N\mathcal{O}(z_n,\bar{z}_n)\rangle\ ,\nonumber
\ea
where we used conformal integrals of Ref.\cite{Osborn1108}, see Appendix B, in particular Eq.(\ref{eq:3confInt}).\footnote{The integrals are divergent as it is obvious from the presence of the $\Gamma(0)$ prefactor. We can regularize the divergence by taking the soft limit on the shadow transform at the end. In other words use  $$ \int \frac{d^2 z_1}{(z-z_1)^{4+i\lambda} (\bz-\bz_1)^{i\lambda} }$$
and take $\lambda\to 0$ at the end. Our results will be the same } The remaining terms can be written as
\ba
-2h_i\partial_w\int&&\!\!\!\!\!\!\frac{d^2 z_1}{Z^2W}\frac{z_1-z_i}{(\bar{z}_1-\bar{z}_i)
(\bar{z}_1-\bar{z})}(\bar{z}_i-\bar{z})\langle\prod_{n=2}^N
\mathcal{O}(z_n,\bar{z}_n)\rangle\\
&=&2h_i\Gamma(0)\left( \frac{1}{(z-w)^2}+2\frac{w-z_i}{(z-w)^3}\right)
\langle\prod_{n=2}^N\mathcal{O}
(z_n,\bar{z}_n)\rangle\ .
\ea
In this way,
the last term of Eq.(\ref{step31}) becomes
\ba
&&2\Gamma(0)\left[ -\frac{1}{(z-w)^3}\sum_{i=2}^N[ (z_i-w)^2\partial_i +2h_i(z_i-w)]\langle\prod_{n=2}^N\mathcal{O}
(z_n,\bar{z}_n)\rangle\right]\\
&&~~+2\Gamma(0)\left[\frac{1}{(z-w)^2}\sum_{i=2}^N[ (z_i-w)\partial_i+h_i]\langle\prod_{n=2}^N\mathcal{O}
(z_n,\bar{z}_n)\rangle\right] = 0\ ,
\ea
with both terms vanishing separately as a consequence of global conformal (Lorentz) invariance \cite{Stieberger:2018onx}:
\ba\nonumber
\sum_{i=2}^N \partial_i\langle\prod_{n=2}^N\mathcal{O}
(z_n,\bar{z}_n)\rangle&=&\sum_{i=2}^N \left( z_i \partial_i+ h_i\right) \langle\prod_{n=2}^N\mathcal{O}
(z_n,\bar{z}_n)\rangle=\\[1mm] \nonumber &=&\sum_{i=2}^N \left( z_i ^2\partial_i+ 2 z_i h_i\right)\langle\prod_{n=2}^N\mathcal{O}
(z_n,\bar{z}_n)\rangle =0\ .\ea
The final result is the expected $T(w)T(z)$ OPE:
\begin{equation}
T(w)T(z) = \label{ttfin} \frac{2T(z)}{(w-z)^2}+\frac{\partial_zT(z)}{w-z}+\makebox{regular}\ .
\end{equation}
\subsection{$\boldmath{T\overline{T}}$}
By repeating the steps leading to Eq.(\ref{eq:TTstep1}), we obtain
\ba\label{eq:TTstep11}
\langle T(w) \overline T(z)&&\!\!\!\!\prod_{n=2}^N\mathcal{O}(z_n,\bar{z}_n)\rangle =\\&&
=\lim_{\Delta_1\rightarrow0}\frac{3!}{2\pi}\int \frac{d^2 z_1}{(\bz-\bz_1)^4}\left( \frac{1}{w-z_1}\partial_{z_1}+\frac{1}{(w-z_1)^2}\right)\nonumber
{\cal G}^+(z_1,\dots)\ , \ea
where \be{\cal G}^+(z_1,\dots)=\langle\label{gpll}
\cO^+_{\Delta_1,+2 }(z_1,\bz_1)\prod_{n=2}^N\mathcal{O}(z_n,\bar{z}_n)\rangle\ ,
\ee
which can be simplified to
\be\label{eq:TTstep12}
\langle T(w) \overline T(z)\prod_{n=2}^N\mathcal{O}(z_n,\bar{z}_n)\rangle
=-\lim_{\Delta_1\rightarrow0}\frac{3!}{2\pi} \int \frac{d^2 z_1}{w-z_1}{\cal G}^+(z_1,\dots)\partial_{z_1}(\bz-\bz_1)^{-4}\nonumber
\ . \ee
After using
\begin{equation}
\partial_{z_1}(\bar{z}-\bar{z}_1)^{-4} =\frac{1}{3!}\partial_{z_1}\bar{\partial}^3_{\bar{z}_1}\frac{1}{\bar{z}-\bar{z}_1} =\frac{2\pi}{3!} \bar{\partial}_{\bar{z}_1}^3\delta^{(2)}(z-z_1)\ ,
\end{equation}
we obtain
\ba
\langle T(w) \overline T(z)\prod_{n=2}^N\mathcal{O}(z_n,\bar{z}_n)\rangle\nonumber
&=&-\lim_{\Delta_1\rightarrow0} \int \frac{d^2 z_1}{w-z_1}{\cal G}^+(z_1,\dots)\bar{\partial}_{\bar{z}_1}^3\delta^{(2)}(z-z_1)\\
&=&\lim_{\Delta_1\rightarrow0}\, \bar{\partial}_{\bar{z}}^3\bigg[\frac{{\cal G}^+(z,\dots)}{w-z}\bigg].
\label{ttbar2}\ea
Note that in  the soft limit
\begin{equation}\lim_{\Delta_1\rightarrow0} {\cal G}^+(z,\dots)= - \frac{1}{2}\sum_{i=2}^N\frac{\bz-\bz_i}{{z}-{z}_i}\frac{{\sigma}-
{z}_i}{{\sigma}-{z}}\big[(\bz-\bz_i)\partial_i -2h_i\big]\langle\prod_{n=2}^N\mathcal{O}(z_n,\bar{z}_n)\rangle\ ,\ee
therefore the derivative $\bar{\partial}_{\bar{z}}^3$ acting on the r.h.s.\ of Eq.(\ref{ttbar2}) gives zero modulo delta-function terms localized at $z=w$ and $z=z_n$. These terms do not affect the Virasoro algebra following from Eq.(\ref{ttfin}). We conclude that up to such delta functions,
\be\label{ttbarfin}
T(w)\overline{T}(z) = \makebox{regular}\ .\ee
\subsection{$TP$ and $\overline TP$}
Recall (\ref{pdef}) that the supertranslation current is defined as
\be
P(z)= \partial_{\bz} \mathcal{O}_{1,+2}(z, \bz)\ .
\ee
The graviton primary field operator $\mathcal{O}_{1,+2}$ has chiral weights
$(\frac{3}{2},-\frac{1}{2})$. We start from
\ba
\langle T(w) P(z)\prod_{n=2}^{N}\mathcal{O}(z_n,\bar{z}_n) \nonumber \rangle&=& \lim_{\Delta_0\rightarrow0}\lim_{\Delta_1\rightarrow1}
\Big(\prod\limits_{n=0}^{N}c(\Delta_n)\Big)\frac{3!}{2\pi}\\ &&\times
\int d^2z_0\frac{1}{(z_0-w)^4}
\partial_{\bz}
\cA_{\ell_0=-2,\ell_1=+2,\ell_2\dots \ell_N}\ .\label{pinteg}
\ea
In the $\Delta_0 \to 0$ limit, c.f.\ Eq.({\ref{mason6}) with $\ell_1=+2$ and the reference point $\xi\to\infty$,
\ba
\partial_{\bz} \lim_{\Delta_0\rightarrow 0} &&\!\!\!\!\!\!\!\Big(\prod\limits_{n=0}^{N}c(\Delta_n)\Big) \cA_{\ell_0=-2,\ell_1=+2,\ell_2\dots \ell_N}\nonumber  \\
&=& -\frac{1}{2} \partial_{\bz} \frac{(z_0-z)}{(\bar{z}_0-\bar{z})}  [ (z_0-z)\partial_{z}-2h_1 ] {\cal G}^+(z,z_2,\dots)\nonumber\\
&&-\frac{1}{2} \partial_{\bz}\sum_{i=2}^N \frac{(z_0-z_{i})}{(\bar{z}_0-\bar{z}_{i})}  [ (z_0-z_i)\partial_{z_i}-2h_i ] {\cal G}^+(z,z_2,\dots) \ ,\label{mason21}
\ea
with ${\cal G}^+$ defined in Eq.(\ref{gpll}) and $h_1=\frac{3}{2}$.
As in previous cases, the sum over $i\ge 2$ is non-singular at $z=w$, therefore we are left with
\ba
\partial_{\bz} &&\!\!\!\!\!\!\!\!\!\lim_{\Delta_0\rightarrow 0} \Big(\prod\limits_{n=0}^{N}c(\Delta_n)\Big) \cA_{\ell_0=-2,\ell_1=+2,\ell_2\dots \ell_N} \\ \nonumber
&=&\frac{1}{2}\Big[3\frac{z_0-z}{(\bar{z}_0-\bar{z})^2} -\frac{(z_0-z)^2}{(\bar{z}_0-\bar{z})^2}\partial_z +3\frac{z_0-z}{\bar{z}_0-\bar{z}}\partial_{\bar{z}} -\frac{(z_0-z)^2}{(\bar{z}_0-\bar{z})}\partial_z\partial_{\bar{z}} \Big]{\cal G}^+(z,z_2,\dots)+\dots
\ea
Upon inserting the above expression into the shadow integral of Eq.(\ref{pinteg}), the first two terms give rise to  delta-function singularities at $z=w$, but vanish for $z\neq w$. In particular,
\be
3\int \frac{d^2 z_0}{(z_0-w)^4}\frac{z_0-z}{(\bar{z}_0-\bar{z})^2} = \partial_w\int \frac{d^2 z_0}{(z_0-w)^3}\frac{z_0-z}{(\bar{z}_0-\bar{z})^2}\sim
\partial_w \delta^{(2)}(z-w) \ .
\ee
where we used conformal integrals (\ref{eq:confint}) and (\ref{eq:2confint}) from Appendix B.
Similarly,
\be
\int \frac{d^2 z_0}{(z_0-w)^4}\frac{(z_0-z)^2}{(\bar{z}_0-\bar{z})^2}
\sim\delta^{(2)}(z-w)
\ee
In this way, we obtain
\ba
\langle T(w)P(z)&& \!\!\!\!\!\!\!\! \prod_{n=2}^{N}\mathcal{O}(z_n,\bar{z}_n)  \rangle=\nonumber \\ \lim_{\Delta_1\rightarrow1}\frac{3}{2\pi}
&&\!\!\!\!\!\!\!\!\int d^2{z_0}\frac{1}{(z_0-w)^4} \Big[3\frac{z_0-z}{\bar{z}_0-\bar{z}}
\partial_{\bar{z}}-\frac{(z_0-z)^2}{(\bar{z}_0-\bar{z})}
\partial_z\partial_{\bar{z}} \Big]  {\cal G}^+(z,z_2,\dots)   +\cdots\\
&=&\frac{3}{2\pi}\int d^2{z_0}\frac{1}{(z_0-w)^4} \Big[3\frac{z_0-z}{\bar{z}_0-\bar{z}}+\frac{(z_0-z)^2}{(\bar{z}_0-\bar{z})}
\partial_z \Big]\langle P(z)\prod_{n=2}^{N}\mathcal{O}(z_n,\bar{z}_n)  \rangle+\cdots\nonumber
\ea
After performing the shadow integrals, we obtain
\be
T(w)P(z) = \frac{3}{2(w-z)^2}P(z)+\frac{1}{w-z}\partial_{z}P(z)+\makebox{regular}\, ,\label{TPope}
\ee
modulo delta-function singularities at $z=w$. Next, we consider
\ba
\langle \overline T(\bw) P(z)\prod_{n=2}^{N}\mathcal{O}(z_n,\bar{z}_n) \nonumber \rangle&=& \lim_{\Delta_0\rightarrow0}\lim_{\Delta_1\rightarrow1}
\Big(\prod\limits_{n=0}^{N}c(\Delta_n)\Big)\frac{3!}{2\pi}\\ &&\times
\int d^2z_0\frac{1}{(\bz_0-\bw)^4}
\partial_{z}
\cA_{\ell_0=+2,\ell_1=+2,\ell_2\dots \ell_N}\ .\label{pinteg}
\ea
After taking the $\Delta_0 \to 0$ limit, we obtain
\ba
\langle\overline T(\bw)P(z)&& \!\!\!\! \prod_{n=2}^{N}\mathcal{O}(z_n,\bar{z}_n)  \rangle=\label{tbatpc} \\ \lim_{\Delta_1\rightarrow1}
\frac{3}{2\pi}
&&\!\!\!\!\int d^2{z_0}\frac{1}{(\bz_0-\bw)^4} \Big[\frac{1}{z_0-z}+\frac{\bz_0-\bz}{{z}_0-{z}}
\partial_{\bz}-\frac{(\bz_0-\bz)^2}{{z}_0-{z}}
\partial_{\bz}^2 \Big]  {\cal G}^+(z,z_2,\dots)   +\dots\nonumber
\ea
where, as in the previous case, we ignored delta-function singularities at $z=w$. As a result,
\be
\overline T(\wbar) {P}(z)=-\frac{\cO_{1,+2}(z,\bar z)}{(\bw-\bz)^3} + \frac{1}{2(\bw-\bz)^2} {P}(z) +\frac{1}{\bw-\bz}\partial_{\zbar} {P}(z)+\makebox{regular},\label{eq:bTPope}\ee
which is the OPE expected for a level one descendant of a primary field.
\section{Extended BMS algebra}\label{sec:BMSalg}
In this section, we discuss connections between the OPEs and extended BMS algebra $\mathfrak{bms_4}$; we also make contact with Ref.\cite{Barnich1703}. The OPEs of the stress energy tensor confirm the presence of Virasoro subalgebra with zero central charge.\footnote{This may change once EYM quantum corrections are taken into account \cite{distler}.}
Here, we concentrate on the role of supertranslations and derive the complete form of the algebra based on the OPEs.

It will be convenient to work with primary fields normalized in a slightly different way. Up to this point, we considered
\be
{\cal O}_{\D, \ell}(z,\bz)= c(\D) \phi^{h,\bh}(z,\bz)\ ,\qquad\textstyle (h, \bar h)=\big(\frac{\Delta+\ell}{2},\frac{\Delta-\ell}{2}\big)\ .
\ee
The correlation functions of  ``bare'' $\phi$ fields are given directly by Mellin-transformed amplitudes:
\be \Big\langle\prod_{n=0}^N\phi^{h_n,\bh_n}(z_n,\bz_n)\Big\rangle={\cal A}_{\ell_1\dots\ell_N}(\Delta_n,z_n,\bz_n)\ee
In the following discussion, the primary field operators will be identified with $\phi^{h,\bh}(z,\bz)$, although the normalization of soft graviton operators $\cO_{0,\pm 2}$
and $\cO_{1,\pm 2}$, and of the related $T$ and $P$ will remain unchanged.

The four-momentum (translation) operators were constructed in Ref.\cite{Stieberger:2018onx}. Anticipating their placement in $\mathfrak{bms_4}$, we define
\be\left(
\begin{array}{cc}
P_{-\ha,-\ha}=P_0+P_3=e^{(\partial_h+ \partial_{\bar{h}})/2}& ~~~P_{-\ha, \ha}=P_1-iP_2= \bz   e^{(\partial_h+ \partial_{\bar{h}})/2}\\[2mm]
P_{\ha,-\ha}=P_1 + i P_2= z  e^{(\partial_h+ \partial_{\bar{h}})/2}&
~~~P_{\ha,\ha}=P_0-P_3=z \bz   e^{(\partial_h+ \partial_{\bar{h}})/2}\end{array}\right),\ee
so that
\be \big[P_{\pm\ha,\pm\ha},\phi^{h,\bh}(z,\bz)\big]=z^{\ha\pm\ha}\bz^{\ha\pm\ha}
\phi^{h+\ha,\bh+\ha}(z,\bz)\label{phalf}\ .\ee

We will use $P_{-\ha,-\ha}$ as a starting point for constructing all supertranslation generators. To that end, we will utilize the extended
BMS algebra \cite{Barnich1703}. In addition to the Virasoro subalgebra, $\mathfrak{bms_4}$ consists of
\begin{equation}
\begin{aligned}\label{eq:BMSalg}
&[P_{ij}, P_{kl}]=0\ ,\\
&[L_n, P_{kl}]= \big({1\over 2} n-k\big)P_{n+k,l}+ n(n^2-1) C_{n,k}\ ,\\
& [\bar{L}_n, P_{kl}]= ({1\over 2}n-l) P_{k,n+l}+n (n^2-1) \bar{C}_{n,l}\ ,
\end{aligned}
\end{equation}
with $m,n\in \mathbb{Z}$ and $i,j,k,l \in \mathbb{Z}+\ha$. Here $C$ and $\bar C$ are possible central extension operators; we will set $C=\bar C =0$ because they cannot be determined by using our methods. The subset $P_{n-\ha,-\ha}$ can be generated in the following way:
\be P_{n-\ha,-\ha}=\frac{1}{i\pi(n+1)}\oint dw\, w^{n+1}\big[T(w),P_{-\ha,-\ha}\big]\ .\label{pkha}
\ee
In order to find out how these operators act on a generic primary field, we first determine the OPE of $\big[T(w),P_{-\ha,-\ha}\big]$  and $\phi^{h,\bh}(z,\bz)$.
We use
 \be [P_{-\ha,-\ha},\cO_{0,-2}(z,\bz)]=-2\cO_{1,-2}(z,\bz)\ ,\ee
 where the factor ${-}2$ is coming from the ratio of normalization factors, $f(0)/f(1)=-2$, and take the $\Delta=1$ soft limit as in Eq.(\ref{mason2}), to show that
\be\label{eq:tPalg3}
[\cO_{0,-2}(w_0,\bw_0), P_{-\ha,-\ha}]\phi^{h_i, \bh_i}(z_i, \bz_i)  = \frac{1}{2} {w_0-z_i \over \bw_0-\bz_i} \phi^{h_i+\ha, \bh_i+\ha}(z_i,\bz_i)+\makebox{regular}\ .
\ee
After taking the shadow transform, we obtain
\be
\label{eq:tPalg4}
 [T(w), P_{-\ha,-\ha}]  \phi^{h, \bh}(z,\bz) =\ha  {1 \over (w-z)^2} \phi^{h+\ha,\bh+\ha}(z,\bz)+\makebox{regular}\ .
\ee
Finally, after combining Eq.(\ref{pkha}) with the above OPE, we obtain
\be
\big[P_{n-\ha,-\ha},\phi^{h, \bh}(z, \bz)\big]=z^n\phi^{h+\ha, \bh+\ha}(z, \bz)\ ,\ee
which agrees with Eq.(\ref{phalf}) for $n=0,1$ and is in agreement with supertranslations of  primary field operators proposed in Ref.\cite{Barnich1703}.
The remaining $\mathfrak{bms_4}$ generators can be constructed as
\be P_{n-\ha,m-\ha}=\frac{1}{i\pi(m+1)}\oint d\bw\, \bw^{m+1}\big[\bar{T}(\bw),P_{n-\ha,-\ha}\big]\ .\label{pkham}
\ee
By following the same steps as before, is easy to show that
\be
\big[P_{n-\ha,m-\ha},\phi^{h, \bh}(z, \bz)\big]=z^n\bz^m\phi^{h+\ha, \bh+\ha}(z, \bz)\ .\label{pall}\ee

Following Ref.\cite{Barnich1703}, we can combine all supertranslation generators into one primary field operator
\be\label{pfield}
{\cal P}(z,\bz) \equiv \sum\limits_{n,m \in \mathbb{Z}} P_{n-\ha, m-\ha}z^{-n-1} \bz^{-m-1}
\ee
${\cal P}$ is a scalar operator with dimension 3, $(h,\bar h)=(\frac{3}{2},\frac{3}{2})$. The commutators (\ref{pall}) are equivalent to the OPE
\be {\cal P}(w,\bw)\phi^{h, \bh}(z, \bz)=\frac{1}{w-z}\frac{1}{\bw-\bz}\phi^{h+\ha, \bh+\ha}(z, \bz)+\makebox{regular}\ .\ee

At this point, we can make connection to the supertranslation operator $P(z)$ discussed in sections 3.2 and 4.3. For bare fields, the OPEs of Eqs.(\ref{poope1}) and (\ref{poope2}) read
\be P(w)\phi^{h, \bh}(z,\bz) =\frac{1}{4} {1 \over w-z}\phi^{h+\ha, \bh+\ha}(z,\bz)+\makebox{regular}\ .\label{poope3} \ee
If we expand
\be\label{ppfield}
P(z)= \sum\limits_{n \in \mathbb{Z}}\widehat P_{n-\ha}z^{-n-1}
\ee
and compare Eq.(\ref{poope3}) with (\ref{pall}), we find $4\widehat P_{n-\ha}= P_{n-\ha,-\ha}$, therefore
\be P(z)=\frac{1}{8\pi i}\oint d\zbar\,{\cal P}(z,\bz) \ee
and a similar expression for $\bar P(\zbar)$. Note that  both $P(z)$ and $\bar P(\bz)$ miss the ``mixed'' operators $P_{n-\ha, m-\ha}$ with $n\neq 0$ and $m\neq 0$ simultaneously.
This is to be contrasted with the primary conformal field operator ${\cal P}(z,\bz)$ of Eq.(\ref{pfield}) which includes all BMS supertranslations.

The operator $P_{-\ha,-\ha}=P_0+P_3$ plays a special role in our construction of BMS generators. With a primary field expanded as
\be
\phi^{h, \bh}(z,\bz)=\sum\limits_{m,n } \phi^{h,\bh}_{m,n} z^{-m-h} \bz^{-n-\bh}\ee
$P_{-\ha,-\ha}$ shifts the modes as:
\be \big[P_{-\ha,-\ha}, \label{shift} \phi^{h,\bh}_{m,n}\big]=\phi^{h+\ha,\bh+\ha}_{m-\ha,n-\ha}\ . \ee

Four-dimensional Minkowski space can be foliated by using Euclidean $AdS_3$ slices \cite{deBoer:2003vf,Cheung:2016iub}. On the other hand it is known that string theory on $AdS_3$ has a spectrum generated by using a spectral flow transformation on the standard representations of $ SL(2;\mathbb{R})$. This transformation shifts conformal dimensions and the mode expansions of primary fields. The transformation in (\ref{shift}) is akin to the spectral flow encountered in string theory on $AdS_3$
\cite{Maldacena:2000hw}.

\section{Conclusions}
Extended BMS is the symmetry of asymptotically flat spacetime and consequently, \linebreak of its hologram on ${\cal C S}^2$. According to Strominger's proposal of flat holography, four-dimensional physics is encoded in two-dimensional CCFT. In particular, Mellin-transformed four-dimensional scattering amplitudes are equivalent to the correlators of primary field operators on ${\cal C S}^2$. In this work, we followed this link. We investigated
the structure of OPEs of the operators generating BMS transformations by using the relation of the collinear limit to the limit of coinciding  points on ${\cal C S}^2$ and the relation of the usual soft limit to the conformally soft limit of operators with integer dimensions. We established a connection between these OPEs and the extended BMS algebra $\mathfrak{bms_4}$. We also showed how the OPEs of the operators representing gauge bosons and gravitons follow directly from the Feynman diagrams of Einstein-Yang-Mills theory.

BMS superrotations are implemented in a rather straightforward way as diffeomorphisms of ${\cal C S}^2$. They are generated by the energy-momentum tensor $T(z)$ given by a shadow transform of the soft graviton operator with dimension 0. Although general shadow transformations are non-local, we found that for soft gravitons, the shadow integrals become localised at operator insertions. As a result, we recovered the standard OPEs which are equivalent to the Virasoro subalgebra without central extension.\footnote{This is also known as Witt algebra.}

BMS supertranslations are more subtle. We investigated the OPEs of the supertranslation operator $P(z)$, a dimension 2 descendant of the soft graviton operator with dimension 1. Indeed, the OPEs of this operator generate shifts of primary field dimensions, as expected from translations and supertranslations. In particular, it contains the momentum component $P_0+P_3$ which generates a flow of chiral weights,
$(h,\bar h)\to(h+\ha,\bar h+\ha)$.
By comparing with BMS algebra, we found however that $P(z)$ and $\bar P(\zbar)$ contain only a subset of supertranslations. They miss not only the non-holomorphic $P_0-P_3$ but also an infinite set of  superstranslations. Nevertheless, we were able to generate the full set by commuting $T(z)$ and $\bar T(\zbar)$ with $P_0+P_3$ which we identified with the $P_{-\ha,-\ha}$ spectral flow generator of $\mathfrak{bms_4}$. All supertranslations can be assembled into a single primary field operator ${\cal P}(z,\bz)$ of dimension 3.

There are many directions for future investigations of CCFT. Since the present discussion relies on the properties of tree-level Einstein-Yang-Mills theory, it would be very interesting to see how four-dimensional quantum corrections are implemented in CCFT. But first and foremost, it is very important to understand the spectrum of CCFT. It is clear that superstranslations generate states very different from conformal wave packets of stable particles. This sector, together with the role of the spectral flow, are certainly worth investigating.

\vskip 1mm
\leftline{\noindent{\bf Acknowledgments}}
\vskip 1mm
\noindent
We are grateful to Glen Barnich and Hugh Osborn for very useful and enlightening correspondence. StSt and TT are grateful to Nima Arkani-Hamed, Monica Pate, Ana-Maria Raclariu and Andy Strominger for stimulating conversations.
This material is based in part upon work supported by the National Science Foundation
under Grant Number PHY--1913328.
Any opinions, findings, and conclusions or recommendations
expressed in this material are those of the authors and do not necessarily
reflect the views of the National Science Foundation.
\renewcommand{\thesection}{A}
\setcounter{equation}{0}
\renewcommand{\theequation}{A.\arabic{equation}}
\renewcommand{\thesection}{A}
\setcounter{equation}{0}
\renewcommand{\theequation}{A.\arabic{equation}}
\vskip 3mm
\section{Collinear limits in EYM theory}\label{App:col}
In this appendix, we use the notation and conventions of Ref.\cite{tt}.
In order to give a precise definition of the leading and subleading collinear parts, we need to specify how the collinear limit is reached from a generic kinematic configuration. Let us specify to generic light-like momenta $p_1,\,p_2$ and introduce two light-like vectors $P$ and $r$ such that the momentum spinors decompose as
\cite{Stieberger:2015kia}:
\ba\lambda_{1}=\lambda_P\cos\theta-\epsilon\lambda_r\sin\theta\ ,
&&\tilde\lambda_{1}=\tilde\lambda_P\cos\theta
-\tilde\epsilon\tilde\lambda_r\sin\theta\ ,
\\[1mm]
\lambda_{2}=\lambda_P\sin\theta+\epsilon\lambda_r\cos\theta\ ,
&&\tilde\lambda_{2}=\tilde\lambda_P\sin\theta+
\tilde\epsilon\tilde\lambda_r\cos\theta\ ,
\ea
hence
\ba p_{1}& ={\bf c}^2P-{\bf s c} (\epsilon
\lambda_r\tilde\lambda_P+\tilde\epsilon\lambda_P\tilde\lambda_r)+ \epsilon\tilde\epsilon{\bf s}^2 r\ ,\label{pon}\\[1mm]
p_{2}& ={\bf s}^2P+{\bf sc} (
\epsilon\lambda_r\tilde\lambda_P+\tilde\epsilon\lambda_P\tilde\lambda_r)+
\epsilon\tilde\epsilon{\bf c}^2 r\ ,\label{ptw}
\ea
where
\be {\bf{c}}\equiv\cos\theta=\sqrt{x}~,\qquad\qquad {\bf{s}}\equiv\sin\theta=\sqrt{1-x}\ .\ee
We also have
\be\langle 12\rangle=\epsilon\,\langle Pr\rangle~, \qquad
[12]=\tilde\epsilon\,[Pr]~.\ee
The total momentum is:
\be p_{1}+p_2=P+\epsilon\tilde\epsilon r~,\qquad (p_{1}+p_2)^2= 2 Pr\,\epsilon\tilde\epsilon\ .\ee
The collinear configuration will be reached in the limit of $\epsilon=\tilde\epsilon= 0$.
With the momenta parametrized by celestial sphere coordinates as in Eq.(\ref{momparam}), this limit corresponds to $z_1=z_2$, i.e.
 \be \label{psum}p_1+p_2=P\equiv \omega_Pq_P\ ,\ee
 with
 \be\label{psum1}\om_P=\om_1+\om_2\, \qquad q_P=q_1=q_2~~(z_P=z_1=z_2,~\bz_P=\bz_1=\bz_2)\ ,\ee
 therefore
 \be \frac{\omega_1}{\omega_P}={\bf c}^2\ ,\qquad \frac{\omega_2}{\omega_P}={\bf s}^2\ .\ee

Yang-Mills amplitudes are singular in the collinear limit. They contain $1/\epsilon$ and $1/\tilde\epsilon$ poles while subleading terms are of order 1 \cite{Stieberger:2015kia}. Here, we are interested in gravitational couplings. These contain singular ratios
\be\label{singe}
\frac{\langle 12\rangle}{[12]}=\frac{\epsilon}{\tilde\epsilon}\frac{\langle Pr\rangle}{[Pr]}=-
\frac{z_{12}}{\bz_{12}}~,\qquad
\frac{[12]}{\langle 12\rangle}=\frac{\tilde\epsilon}{\epsilon}\frac{[Pr]}{\langle Pr\rangle}=-
\frac{\bz_{12}}{z_{12}}\ .\ee
We will extract them from EYM Feynman diagrams. To that end, it is convenient to rewrite Eqs.(\ref{pon},\ref{ptw}) as
\ba p_{1}&=&{\bf c}^2P-\frac{\bf s c}{\sqrt 2} \nonumber \big(\epsilon\epsilon_P^+\langle rP\rangle
+\tilde\epsilon\epsilon_P^-[Pr]\big) +  \epsilon\tilde\epsilon{\bf s}^2 r\\ &&={\bf c}^2P+\frac{\bf s c}{\sqrt 2} \big(\epsilon_P^+\langle 12\rangle \label{p1col}
-\epsilon_P^-[12]\big)+ \epsilon\tilde\epsilon{\bf s}^2 r\\[1mm]
p_{2}&=&{\bf s}^2P+\frac{\bf s c}{\sqrt 2} \nonumber \big(\epsilon\epsilon_P^+\langle rP\rangle
+\tilde\epsilon\epsilon_P^-[Pr]\big)+ \epsilon\tilde\epsilon{\bf c}^2 r\\ && ={\bf s}^2P-\frac{\bf s c}{\sqrt 2} \big(\epsilon_P^+\langle 12\rangle
-\epsilon_P^-[12]\big)+ \epsilon\tilde\epsilon{\bf c}^2 r
\ea
Here, $\epsilon_P^\pm$ are the  polarization vectors of a massless vector boson with momentum $P$ with the gauge reference vector $r$. Note that
\ba
\epsilon_1^+ && =\epsilon_P^+ -\sqrt{2}\tilde\epsilon r\frac{\bf s}{{\bf c}\langle rP\rangle}\ , \quad
\epsilon_1^-=\epsilon_P^- +\sqrt{2}\epsilon r\frac{\bf s}{{\bf c}[ rP]}\ ,\\
\epsilon_2^+ && =\epsilon_P^+ +\sqrt{2}\tilde\epsilon r\frac{\bf c}{{\bf s}\langle rP\rangle}\ , \quad
\epsilon_2^-=\epsilon_P^- -\sqrt{2}\epsilon r\frac{\bf c}{{\bf s}[ rP]}\ .
\ea
For all polarization vectors labeled by $i=1,2,P$, \be\label{epspro}
\epsilon_i^\pm\cdot\epsilon_j^\mp=-1 \ ,\qquad
\epsilon_i^\pm\cdot\epsilon_j^\pm=\epsilon_i^\pm\cdot r=\epsilon_i\cdot p_i=0.\ee
Other scalar products are also easy to obtain:
\ba\epsilon_1^+\cdot p_2=-\frac{1}{\sqrt{2}}\frac{[12]\bf s}{\bf c}\ , &&\quad \epsilon_2^+\cdot p_1=\frac{1}{\sqrt{2}}\frac{[12]\bf c}{\bf s}\ ,\label{spr1}\\
\epsilon_1^-\cdot p_2=\frac{1}{\sqrt{2}}\frac{\langle 12\rangle\bf s}{\bf c}\ , &&\quad \epsilon_2^-\cdot p_1=-\frac{1}{\sqrt{2}}\frac{\langle 12\rangle\bf c}{\bf s}\ .\label{spr2}\ea

The singularities (\ref{singe}) appear only in Feynman diagrams involving ``splitting'' of a massless virtual particle into the collinear pair.  They originate from the propagator poles $1/(p_1+p_2)^2\sim (\epsilon\tilde\epsilon)^{-1}$. The vertices describing gravitational splitting  ``soften'' these poles
by terms which are quadratic in $\epsilon$ and $\tilde\epsilon$. Diagrams without such splitting are non-singular, therefore collinear singularities have a universal form that can be extracted from three-point vertices. In our discussion, we will be using the graviton polarization tensors
\be \epsilon^{\mu\nu}(\ell=\pm 2)=\epsilon^{\pm\mu}\epsilon^{\pm\nu}\ .\label{grapol}\ee

\subsection{Collinear gravitons}\label{sec:colgrav}
EYM amplitudes with external gravitons are singular when two gravitons become collinear. Feynman diagrams contributing to collinear singularities must necessarily contain a virtual graviton splitting into two external gravitons, as shown in Fig. \ref{3gvertex}.
\begin{figure}
\begin{center}
\begin{picture}(300,100)(0,-10)
\SetColor{Black}
\Text(100,90)[l]{$\alpha\beta$}
\Text(200,90)[l]{$\gamma\delta$}
\Text(120,30)[l]{$\mu\nu$}
\Text(120,50)[l]{$\rho\tau$}
\Photon(150,20)(150,60){5}{4}\Vertex(150,60){2}
\Photon(152,20)(152,60){5}{4}
\ArrowLine(170,30)(170,48)
\Text(175,38)[l]{$p_1+p_2$}
\Photon(150,60)(120,90){5}{4}
\Photon(151.5,61.5)(121.5,91.5){5}{4}
\ArrowLine(120,65)(104,80)
\Text(95,70)[l]{$p_1$}
\Photon(150,60)(180,90){5}{4}
\Photon(148.5,61.5)(178.5,91.5){5}{4}
\ArrowLine(175,70)(190,85)
\Text(190,70)[l]{$p_2$}
\GCirc(150,0){22}{0.5}
\Vertex(125,18){1}
\Vertex(122.5,10.5){1}
\Vertex(120,3){1}
\Vertex(175,18){1}
\Vertex(178.5,10.5){1}
\Vertex(180,3){1}
\Vertex(135,-26){1}
\Vertex(150,-26){1}
\Vertex(165,-26){1}
\end{picture}
\end{center}
\caption{Feynman diagrams leading to collinear graviton singularities.}\label{3gvertex}
\end{figure}
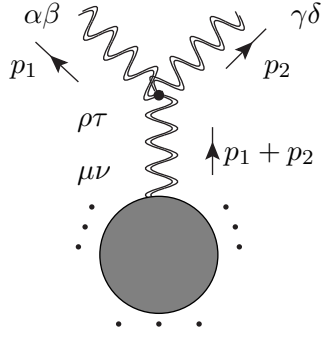
We will be considering the ``splitting'' tensor
\begin{equation}
S^{\mu\nu}= \epsilon_1^{\alpha\beta}\epsilon_2^{\gamma\delta} D^{\mu\nu}_{\quad\rho\tau}(p_1+p_2)V^{\rho\tau}_{\alpha\beta, \gamma\delta}(p_1,p_2)\ ,
\end{equation}
where
$V^{\rho\tau}_{\alpha\beta, \gamma\delta}(p_1,p_2)$ is the
three-graviton vertex
three-graviton vertex
\ba
&&V_{\alpha\beta,\gamma\delta,\rho\tau}(p_1,p_2)=
 i\kappa\Bigg\{ -\frac{1}{2}(p_1^2+p_2^2+(p_1+p_2)^2)\times \bigg[ I^{\sigma}_{~\alpha,\rho\tau}I_{\gamma\delta,\sigma\beta}\nonumber\\
&&~~~~~+\, I^{\sigma}_{~\beta,\rho\tau}
I_{\gamma\delta,\sigma\alpha}+\frac{1}{4}g_{\rho\tau}g_{\alpha\beta}g_{\gamma\delta}
-\frac{1}{2}( g_{\rho\tau}I_{\alpha\beta,\gamma\delta}+g_{\alpha\beta}I_{\rho\tau,\gamma\delta}+
g_{\gamma\delta}I_{\rho\tau,\alpha\beta})\bigg]\label{vthreegrav}\\
&+&p_1^{\sigma}p_2^{\lambda}\bigg[I_{\alpha\beta,\gamma\delta}I_{\rho\tau,\sigma\lambda}-\frac{1}{2}(I_{\rho\tau,\delta\sigma}I_{\alpha\beta,\gamma\lambda}
+I_{\rho\tau,\gamma\sigma}I_{\alpha\beta,\delta\lambda}+I_{\gamma\delta,\beta\sigma}I_{\rho\tau,\alpha\lambda}+I_{\gamma\delta,\alpha\sigma}I_{\rho\tau,\beta\lambda})\bigg]\nonumber\\
&-&(p_1+p_2)^{\sigma}p_2^{\lambda}\bigg[I_{\rho\tau,\gamma\delta}I_{\alpha\beta,\sigma\lambda}-\frac{1}{2}(I_{\alpha\beta,\delta\sigma}I_{\rho\tau,\gamma\lambda}+I_{\alpha\beta,\gamma\sigma}I_{\rho\tau,\delta\lambda}+I_{\gamma\delta,\rho\sigma}I_{\alpha\beta,\tau\lambda}+I_{\gamma\delta,\tau\sigma}I_{\alpha\beta,\rho\lambda})\bigg]\nonumber\\
&-&(p_1+p_2)^{\sigma}p_1^{\lambda}\bigg[I_{\rho\tau,\alpha\beta}I_{\gamma\delta,\sigma\lambda}
-\frac{1}{2}(I_{\gamma\delta,\tau\lambda}I_{\alpha\beta,\rho\sigma}+I_{\gamma\delta,\rho\lambda}
I_{\alpha\beta,\tau\sigma}+I_{\gamma\delta,\beta\sigma}I_{\rho\tau,\alpha\lambda}+
I_{\gamma\delta,\alpha\sigma}I_{\rho\tau,\beta\lambda})\bigg]\Bigg\},\nonumber\ea
with
\be
I_{\rho\tau,\sigma\lambda}= \frac{1}{2}(g_{\rho\sigma}g_{\tau\lambda}+g_{\rho\lambda}g_{\tau\sigma})\ ,\label{3gravperinv}
\ee
see  Refs.\cite{DeWitt1967} and \cite{Bohr2004}.
For the graviton propagator, we use de Donder gauge with
\begin{equation}
D^{\mu\nu}_{\quad\rho\tau}(p_1+p_2)  = \frac{i}{2}\left( \delta^{\mu}_{\rho}\delta^{\nu}_{\tau}+\delta^{\mu}_{\tau}
\delta^{\nu}_{\rho}-g^{\mu\nu}
g_{\rho\tau}\right)\frac{1}{(p_1+p_2)^2}
\end{equation}

We first consider the (simpler) case of identical helicities with the
polarization tensors
\begin{equation}
\epsilon_1^{\alpha\beta}(\ell=+2)=\epsilon_1^{+\alpha}\epsilon_1^{+\beta}, \qquad \epsilon_2^{\gamma\delta}(\ell=+2)=\epsilon_2^{+\gamma}\epsilon_2^{+\delta}\ .
\end{equation}
In the corresponding splitting tensor, most of terms disappear due to Eqs.(\ref{epspro}). As result,
\be\label{eq:I+2+2}\nonumber
S^{\mu\nu}(+2,+2)= \frac{\kappa}{(p_1+p_2)^2}\big(\epsilon_2^+ \cdot p_1\, \epsilon_2^+\cdot p_1\, \epsilon_1^{+\mu}\epsilon_1^{+\nu}-\epsilon_1^+\cdot p_2\,\epsilon_2^+\cdot p_1\, \epsilon_1^{+\mu}\epsilon_2^{+\nu}\big)+(1\leftrightarrow 2)\ .
\ee
By using Eqs.(\ref{p1col}-\ref{spr2}) and neglecting terms of order $O(\tilde\epsilon^2)$, we obtain
\be S^{\mu\nu}(+2,+2)=
-\frac{\kappa}{2}\frac{[12]}{\langle 12\rangle}\frac{1}{{\bf c}^2{\bf s}^2}\epsilon_P^{+\mu}\epsilon_P^{+\nu}=\frac{\kappa}{2}
\frac{\omega_P^2}{\omega_1\omega_2}\frac{\bar{z}_{12}}{z_{12}}\,
\epsilon_P^{+\mu}\epsilon_P^{+\nu}
\ ,
\ee
which leads to Eq.(\ref{twogr}) after setting $\kappa=2$. The result agrees with Ref.\cite{Bern1998}.

The case of opposite helicities, with
\begin{equation}
\epsilon_1^{\alpha\beta}(\ell=-2)=\epsilon_1^{-\alpha}\epsilon_1^{-\beta}, \qquad \epsilon_2^{\gamma\delta}(\ell=+2)=\epsilon_2^{+\gamma}\epsilon_2^{+\delta}\ .
\end{equation}
is a little bit more complicated. The corresponding splitting tensor reads
\ba
S^{\mu\nu}(-2,+2)
&=&\frac{\kappa}{ (p_1+ p_2)^2}\bigg[ (\epsilon_1^- \cdot p_2)^2 \epsilon_2^{+\mu}\epsilon_2^{+\nu}-\frac{1}{2} (\epsilon^-_1\cdot\epsilon^+_2)^2p_1^{\mu}\,p_2^{\nu} \nonumber \\
 &+&\frac{1}{2}(\epsilon_1^-\cdot \epsilon_2^+)(\epsilon_1^-\cdot p_2)p_1^{\mu}\epsilon_2^{+\nu} +\frac{1}{2}(\epsilon_1^-\cdot \epsilon_2^+)(\epsilon_1^-\cdot p_2)p_1^{\nu}\epsilon_2^{+\mu} \nonumber\\
&-&\frac{1}{2} (\epsilon_1^-\cdot \epsilon_2^+)(\epsilon_2^+\cdot p_1)p_1^{\mu}\epsilon_1^{-\nu} -\frac{1}{2} (\epsilon_1^-\cdot \epsilon_2^+)(\epsilon_2^+\cdot p_1)p_1^{\nu}\epsilon_1^{-\mu} \nonumber\\
&-&\epsilon_1^-\cdot p_2\,\epsilon_2^+\cdot p_1\, \epsilon_1^{-\mu}\epsilon_2^{+\nu}+\frac{1}{2}(p_1+p_2)^2\epsilon_1^-\cdot \epsilon_2^+\epsilon_1^{-\mu}\epsilon_2^{+\nu}\bigg]\nonumber\\
&+&(1,-\leftrightarrow 2,+) \ .
\ea
We are interested in terms with $\epsilon^{-1}$ or $\tilde\epsilon^{-1}$ poles only. These come from
\ba
S^{\mu\nu}(-2,+2) &=&- \frac{\kappa}{2}\frac{{\bf s}^6}{{\bf c}^2}\frac{\langle 12\rangle}{[12]}(\epsilon_P^+)^{\mu}(\epsilon_P^+)^{\nu}-\frac{\kappa}{2}
\frac{{\bf c}^6}{{\bf s}^2}\frac{[12]}{\langle12\rangle}(\epsilon_P^-)^{\mu}
(\epsilon_P^-)^{\nu}\nonumber +\dots\\[1mm]
&=&\frac{\kappa}{2}\frac{\omega_2^3}{\omega_P^2\omega_1}\frac{z_{12}}{\bar{z}_{12}}
(\epsilon_P^+)^{\mu}(\epsilon_P^+)+\frac{\kappa}{2}\frac{\omega_1^3}
{\omega_P^2\omega_2}\frac{\bar{z}_{12}}{z_{12} }(\epsilon_P^-)^{\mu}
(\epsilon_P^-)^{\nu}\ ,
\ea
which leads to Eq(\ref{clgrav2}) after setting $\kappa=2$.

\subsection{Collinear graviton and gauge boson}\label{sec:colgauge}
The amplitudes involving gauge bosons and graviton are singular in the limit when the graviton becomes collinear with the gauge bosons. In singular Feynman diagrams, virtual gauge bosons radiate gravitons,
as shown in Fig. \ref{2gra1gvertex}.
We define the splitting vector
\begin{equation}
S^{\mu}= \epsilon_1^{\beta}\epsilon_{2\,\gamma\delta} D^{\mu\alpha}(p_1+p_2)
V_{\alpha,\beta}^{\gamma\delta}(p_1,p_2)\ ,
\end{equation}
where $V_{\alpha,\beta}^{\gamma\delta}(p_1,p_2)$ is the EYM vertex which couples two gauge bosons (with identical gauge group indices) to the graviton:
\ba
V_{\alpha,\beta}^{\gamma\delta}(p_1,p_2)&=&\frac{i\kappa}{2}\bigg[ g_{\alpha\beta}[(p_1+p_2)^{\gamma}p_1^{\delta}+
(p_1+p_2)^{\delta}p_1^{\gamma}]-(p_{1}+p_2)_{\beta}
(p_1^{\gamma}\delta^{\delta}_{\alpha}+p_1^{\delta}\delta^{\gamma}_{\alpha} )\nonumber \\
&-&p_{1\alpha}[(p_1+p_2)^{\gamma}\delta_{\beta}^{\delta}+
(p_1+p_2)^{\delta}\delta_{\beta}^{\gamma}]+p_1\cdot (p_1+p_2)(\delta_{\alpha}^{\gamma}\delta_{\beta}^{\delta}+
\delta_{\alpha}^{\delta}\delta_{\beta}^{\gamma})\nonumber \\
&-&g^{\gamma\delta}[ (p_1\cdot (p_1+p_2)g_{\alpha\beta}-p_{1\alpha}(p_1+p_2)_{\beta}]\bigg]\label{vert11}
\ea
and the gauge boson propagator
\begin{equation}
D^{\mu\alpha}(p_1+p_2) = \frac{-i}{(p_1+p_2)^2}g^{\mu\alpha}
\end{equation}
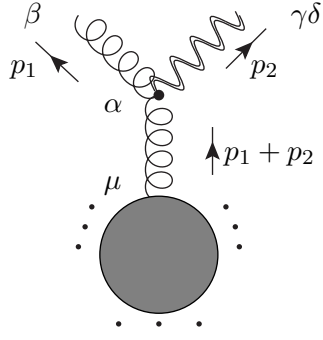
\begin{figure}[h]
\begin{center}
\begin{picture}(300,100)(0,0)
\SetColor{Black}
\Text(100,100)[l]{$\beta$}
\Text(200,100)[l]{$\gamma\delta$}
\Text(130,66)[l]{$\alpha$}
\Text(130,36)[l]{$\mu$}
\Gluon(150,30)(150,70){5}{4}\Vertex(150,70){2}
\ArrowLine(170,40)(170,58)
\Text(175,48)[l]{$p_1+p_2$}
\Gluon(150,70)(120,100){5}{4}
\ArrowLine(120,75)(104,90)
\Text(95,80)[l]{$p_1$}
\Photon(150,70)(180,100){5}{4}
\Photon(148.5,71.5)(178.5,101.5){5}{4}
\ArrowLine(175,80)(190,95)
\Text(185,80)[l]{$p_2$}
\GCirc(150,10){22}{0.5}
\Vertex(125,28){1}
\Vertex(122.5,20.5){1}
\Vertex(120,13){1}
\Vertex(175,28){1}
\Vertex(178.5,20.5){1}
\Vertex(180,13){1}
\Vertex(135,-16){1}
\Vertex(150,-16){1}
\Vertex(165,-16){1}
\end{picture}
\end{center}
\caption{EYM Feynman diagrams leading to singularities for collinear gauge bosons and \newline \hspace*{17mm}gravitons.}\label{2gra1gvertex}
\end{figure}

We first consider the case of all-plus helicities:
\begin{equation}
\epsilon_1^{\beta}(\ell=+1)=\epsilon_1^{+\beta}, \qquad \epsilon_2^{\gamma\delta}(\ell=+2)=\epsilon_2^{+\gamma}\epsilon_2^{+\delta}\ .
\end{equation}
The corresponding splitting vector reads
\ba
S^{\mu}(+2,+1)& =&\nonumber
\frac{\kappa}{(p_1+ p_2)^2} \big[- \epsilon_2^+ \cdot p_1\, \epsilon_1^+ \cdot p_2 \epsilon_1^{+\mu} + (\epsilon_1^+\cdot p_2)^2 \epsilon_2^{+\mu}\big]\\
& =& -\frac{\kappa}{2}\frac{1}{{\bf c}^2}\frac{[12]}{\langle 12\rangle} \epsilon_P^{+\mu}=\frac{\bz_{12}}{z_{12}} \frac{\omega_P}{\omega_1}
\epsilon_P^{+\mu}.\label{spl11}\ea
{}For opposite helicities, we obtain
\ba
S^{\mu}(-2,+1)
 &=&\frac{\kappa}{(p_1+ p_2)^2} \big[- \epsilon_2^+ \cdot p_1\, \epsilon_1^- \cdot p_2 \epsilon_1^{-\mu} + (\epsilon_1^-\cdot p_2)^2 \epsilon_2^{+\mu}-\epsilon_1^-\cdot \epsilon_2^+\, \epsilon_1^- \cdot p_2\, p_2^{\mu}\nonumber \\ && +\,p_1\cdot p_2\epsilon_1^-\cdot \epsilon_2^+ \epsilon_1^{-\mu}\big]= -\frac{\kappa}{2}\frac{\langle 12\rangle}{[12]}\frac{{\bf s}^4}{{\bf c}^2} \epsilon_P^{+\mu}+\cdots
=\frac{\kappa}{2}\frac{z_{12}}{\bar{z}_{12}}\frac{\omega_2^2}{\omega_1 \omega_P} \epsilon_P^{+\mu}.\label{spl12}\ea
where we omitted terms without $\epsilon^{-1}$ or $\tilde\epsilon^{-1}$
poles.  Eqs.(\ref{spl12}) and (\ref{spl12}) lead to Eqs.(\ref{spp11}) and (\ref{spp12}), respectively.
\subsection{Gravitational corrections to the collinear limit of two gauge bosons}
A virtual graviton can split into two gauge bosons, as in Feynman diagrams shown in Fig.\ref{2g1gravertex}. In the collinear limit, the corresponding contributions are ``softer'' than single
$\epsilon^{-1}$ or $\tilde\epsilon^{-1}$ poles due to Yang-Mills interactions \cite{tt}. They contain the same type of $\epsilon/\tilde\epsilon$ and $\tilde\epsilon/\epsilon$ singularities as the gravitational interactions discussed in part {\bf{A.1}}.
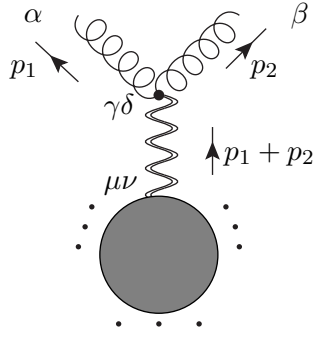
\begin{figure}[h]
\begin{center}
\begin{picture}(300,100)(0,0)
\SetColor{Black}
\Text(100,100)[l]{$\alpha$}
\Text(200,100)[l]{$\beta$}
\Text(130,66)[l]{$\gamma\delta$}
\Text(130,36)[l]{$\mu\nu$}
\Photon(150,30)(150,70){5}{4}\Vertex(150,70){2}
\Photon(152,30)(152,70){5}{4}
\ArrowLine(170,40)(170,58)
\Text(175,48)[l]{$p_1+p_2$}
\Gluon(150,70)(120,100){5}{4}
\ArrowLine(120,75)(104,90)
\Text(95,80)[l]{$p_1$}
\Gluon(150,70)(180,100){5}{4}
\ArrowLine(175,80)(190,95)
\Text(185,80)[l]{$p_2$}
\GCirc(150,10){22}{0.5}
\Vertex(125,28){1}
\Vertex(122.5,20.5){1}
\Vertex(120,13){1}
\Vertex(175,28){1}
\Vertex(178.5,20.5){1}
\Vertex(180,13){1}
\Vertex(135,-16){1}
\Vertex(150,-16){1}
\Vertex(165,-16){1}
\end{picture}
\end{center}
\caption{EYM Feynman diagrams leading to gravitational corrections to Yang-Mills collinear\newline \hspace*{16mm} singularities. }\label{2g1gravertex}
\end{figure}

The gravitational splitting tensor of two gauge bosons is defined as
\be
S^{\mu\nu}_g = \epsilon_1^{\alpha}\epsilon_2^{\beta}V_{\alpha,\beta}^{\gamma\delta}
(p_1,p_2)D^{\mu\nu}_{\quad\gamma\delta}(p_1+p_2)
\end{equation}
with the vertex related by crossing to Eq.(\ref{vert11}):
\ba
V_{\alpha,\beta}^{\gamma\delta}(p_1,p_2)&=&-\frac{i\kappa}{2}\bigg[ -g_{\alpha\beta}(p_1^{\gamma}p_2^{\delta}+p_1^{\delta}p_2^{\gamma})
+p_{1\beta}(p_2^{\gamma}\delta^{\delta}_{\alpha}+p_2^{\delta}
\delta^{\gamma}_{\alpha} )
+p_{2\alpha}(p_1^{\gamma}\delta_{\beta}^{\delta}
+p_1^{\delta}\delta_{\beta}^{\gamma})\nonumber\\ && -p_1\cdot p_2(\delta_{\alpha}^{\gamma}\delta_{\beta}^{\delta}+
\delta_{\alpha}^{\delta}\delta_{\beta}^{\gamma})-g^{\gamma\delta}[ -(p_1\cdot p_2)g_{\alpha\beta}+p_{2\alpha}p_{1\beta}]\bigg].
\ea
It is easy to see that $S^{\mu\nu}_g(+1,+1)$ and $S^{\mu\nu}_g (-1,-1)$
are non-singular. On the other hand,
\ba
S^{\mu\nu}_g(+1,-1)
&=&\frac{\kappa}{2(p_1+p_2)^2}\big[\epsilon_2^-\cdot \epsilon_1^+ p_1^{\mu}p_2^{\nu} +\epsilon_2^-\cdot \epsilon_1^+ p_1^{\nu}p_2^{\mu} - \epsilon_1^+\cdot p_2 p_1^{\mu}\epsilon_2^{-\nu} \nonumber\\
&&-  \epsilon_1^+\cdot p_2 p_1^{\nu}\epsilon_2^{-\mu} -\epsilon_2^-\cdot p_1 p_2^{\mu}\epsilon_1^{+\nu}-\epsilon_2^-\cdot p_1p_2^{\nu}\epsilon_1^{+\mu} \big]\nonumber \\
 &=&\frac{\kappa}{2} \frac{\langle 12 \rangle}{[12]}{\bf c}^4\epsilon_P^{+\mu}\epsilon_P^{+\nu} +\frac{\kappa}{2}\frac{[12]}{\langle 12 \rangle} {\bf s}^4\epsilon_P^{-\mu}\epsilon_P^{-\nu} +\cdots\nonumber\\ &=&
 -\frac{\kappa}{2} \frac{z_{12}}{\bar{z}_{12}}\frac{\omega_1^2}{\omega_P^2}
 \epsilon_P^{+\mu}\epsilon_P^{+\nu} - \frac{\kappa}{2}\frac{\bar{z}_{12}}{z_{12}}
\frac{\omega_2^2}{\omega_P^2}
  \epsilon_P^{-\mu}\epsilon_P^{-\nu}\ .
\ea
After setting $\kappa=2$, the above splitting tensor yields the last two terms of Eq.(\ref{eymcol}).
\renewcommand{\thesection}{B}
\setcounter{equation}{0}
\renewcommand{\theequation}{B.\arabic{equation}}
\renewcommand{\thesection}{B}
\setcounter{equation}{0}
\renewcommand{\theequation}{B.\arabic{equation}}
\vskip 3mm
\section{Conformal Integrals}\label{App:ConfInt}
In the present paper, several integrals over celestial sphere positions have been performed by using the  formulas given in \cite{Osborn1108}.
They have the general form
\be\label{eq:confint}
I_n={1\over \pi} \int\ d^2z\ f_n(z) \bar{f}_n(\bz), \quad f_n(z)= \prod_{i=1}^n {1\over (z-z_i)^{q_i}} \ , \quad \bar{f}_n(\bz)= \prod_{i=1}^n {1\over (\bz-\bz_i)^{\bar{q}_i}}\ ,
\ee
where
\be\label{eq:confw}
\sum_{i=1}^n q_i=\sum_{i=1}^n \bar{q}_i=2 \ , \quad q_i-\bar{q}_i \in \mathbb{Z}\ .
\ee
The two conditions above are necessary for $I_n$ to be covariant under the group $SL(2;\mathbb{C})$ of global conformal transformations and single valued. Convergence of the integral requires $q_i+\bar{q}_i<2$ for all $i$, but we can always extend it by using analytic continuation.
When $n =3$, assuming that all $z_i$ are different, the result is given by
\begin{equation}\label{eq:3confInt}
I_3= K_{123} \, z_{12}^{q_3-1}z_{23}^{q_1-1}z_{31}^{q_2-1}\bar{z}_{12}^{\bar{q}_3-1}
\bar{z}_{23}^{\bar{q}_1-1}\bar{z}_{31}^{\bar{q}_2-1}\ ,
\end{equation}
where the normalization factor
\begin{equation}\label{eq:K123}
K_{123} = \frac{\Gamma(1-q_1)\Gamma(1-q_2)\Gamma(1-q_3)}{\Gamma(\bar{q}_1)
\Gamma(\bar{q}_2)\Gamma(\bar{q}_3)} = \frac{\Gamma(1-\bar{q}_1)\Gamma(1-\bar{q}_2)\Gamma(1-\bar{q}_3)}{\Gamma(q_1)
\Gamma(q_2)\Gamma(q_3)}\ .
\end{equation}
If two $z_i$'s are coincident, the integral becomes
\be\label{eq:2confint}
I_2= K_{12} (-1)^{q_1-\bar{q}_1} \pi \delta^2(q_1-q_2)\ ,
\ee
where
\be\label{eq:K12}
K_{12}= \frac{\Gamma(1-q_1)\Gamma(1-q_2)}{\Gamma(\bar{q}_1)\Gamma(\bar{q}_2)}= \frac{\Gamma(1-\bar{q}_1)\Gamma(1-\bar{q}_2)}{\Gamma(q_1)\Gamma(q_2)} .
\ee


\begin{thebibliography}{99}
\bibitem{Bondi:1962px}
  H.~Bondi, M.~G.~J.~van der Burg and A.~W.~K.~Metzner,
  ``Gravitational waves in general relativity. 7. Waves from axisymmetric isolated systems,''
  Proc.\ Roy.\ Soc.\ Lond.\ A {\bf 269}, 21 (1962).
\bibitem{Sachs:1962wk}
  R.~K.~Sachs,
  ``Gravitational waves in general relativity. 8. Waves in asymptotically flat space-times,''
  Proc.\ Roy.\ Soc.\ Lond.\ A {\bf 270}, 103 (1962).
\bibitem{Barnich:2009se}
  G.~Barnich and C.~Troessaert,
  ``Symmetries of asymptotically flat 4 dimensional spacetimes at null infinity revisited,''
  Phys.\ Rev.\ Lett.\  {\bf 105}, 111103 (2010)
  [arXiv:0909.2617 [gr-qc]].
\bibitem{Strominger:2013jfa}
  A.~Strominger,
  ``On BMS Invariance of Gravitational Scattering,''
  JHEP {\bf 1407}, 152 (2014)
  [arXiv:1312.2229 [hep-th]].
\bibitem{Strominger:2017zoo}
  A.~Strominger,
  {\it Lectures on the Infrared Structure of Gravity and Gauge Theory},
Princeton University Press (2018)
  [arXiv:1703.05448 [hep-th]].

\bibitem{Pasterski:2019msg}
  S.~Pasterski,
  ``Implications of Superrotations,''
  Phys.\ Rept.\  {\bf 829}, 1 (2019)
  [arXiv:1905.10052 [hep-th]].

\bibitem{deBoer:2003vf}
  J.~de Boer and S.~N.~Solodukhin,
  ``A Holographic reduction of Minkowski space-time,''
  Nucl.\ Phys.\ B {\bf 665} (2003) 545 (2003)
  [hep-th/0303006]

\bibitem{Pasterski:2016qvg}
  S.~Pasterski, S.~H.~Shao and A.~Strominger,
  ``Flat Space Amplitudes and Conformal Symmetry of the Celestial Sphere,''
  Phys.\ Rev.\ D {\bf 96} (2017) no. 6, 065026
  [arXiv:1701.00049 [hep-th]].

  \bibitem{Pasterski:2017ylz}
  S.~Pasterski, S.~H.~Shao and A.~Strominger,
  ``Gluon Amplitudes as 2d Conformal Correlators,''
  Phys.\ Rev.\ D {\bf 96} (2017) no. 8, 085006  [arXiv:1706.03917 [hep-th]].

\bibitem{Schreiber:2017jsr}
  A.~Schreiber, A.~Volovich and M.~Zlotnikov,
  ``Tree-level gluon amplitudes on the celestial sphere,''
  Phys.\ Lett.\ B {\bf 781} (2018) 349
  [arXiv:1711.08435 [hep-th]].
\bibitem{Stieberger:2018edy}
  S.~Stieberger and T.~R.~Taylor,
  ``Strings on Celestial Sphere,''
  Nucl.\ Phys.\ B {\bf 935} (2018) 388
  [arXiv:1806.05688 [hep-th]].

\bibitem{Pasterski:2017kqt}
  S.~Pasterski and S.~H.~Shao,
  ``Conformal basis for flat space amplitudes,''
  Phys.\ Rev.\ D {\bf 96} (2017) no. 6, 065022
  [arXiv:1705.01027 [hep-th]].

\bibitem{Fotopoulos:2019tpe}
  A.~Fotopoulos and T.~R.~Taylor,
  ``Primary Fields in Celestial CFT,''
  JHEP {\bf 1910}, 167 (2019)
  [arXiv:1906.10149 [hep-th]].

\bibitem{Donnay:2018neh}
  L.~Donnay, A.~Puhm and A.~Strominger,
  ``Conformally Soft Photons and Gravitons,''
  JHEP {\bf 1901} (2019) 184
  [arXiv:1810.05219 [hep-th]].


\bibitem{FFT2019}
  W.~Fan, A.~Fotopoulos and T.~R.~Taylor,
  ``Soft Limits of Yang-Mills Amplitudes and Conformal Correlators,''
  JHEP {\bf 1905} (2019) 121
  [arXiv:1903.01676 [hep-th]].
\bibitem{Kapec:2016jld}
  D.~Kapec, P.~Mitra, A.~M.~Raclariu and A.~Strominger,
  ``2D Stress Tensor for 4D Gravity,''
  Phys.\ Rev.\ Lett.\  {\bf 119} (2017) no.12,  121601
  [arXiv:1609.00282 [hep-th]].

\bibitem{Cheung:2016iub}
  C.~Cheung, A.~de la Fuente and R.~Sundrum,
  ``4D scattering amplitudes and asymptotic symmetries from 2D CFT,''
  JHEP {\bf 1701} (2017) 112
  [arXiv:1609.00732 [hep-th]].
\bibitem{Pate:2019lpp}
  M.~Pate, A.~M.~Raclariu, A.~Strominger and E.~Y.~Yuan,
  ``Celestial Operator Products of Gluons and Gravitons,''
  arXiv:1910.07424 [hep-th].
\bibitem{Pate:2019mfs}
  M.~Pate, A.~M.~Raclariu and A.~Strominger,
  ``Conformally Soft Theorem in Gauge Theory,''
  Phys.\ Rev.\ D {\bf 100}, no. 8, 085017 (2019)
  [arXiv:1904.10831 [hep-th]].


\bibitem{Nandan:2019jas}
  D.~Nandan, A.~Schreiber, A.~Volovich and M.~Zlotnikov,
  ``Celestial Amplitudes: Conformal Partial Waves and Soft Limits,''
  JHEP {\bf 1910}, 018 (2019)
  [arXiv:1904.10940 [hep-th]].

\bibitem{Adamo:2019ipt}
  T.~Adamo, L.~Mason and A.~Sharma,
  ``Celestial amplitudes and conformal soft theorems,''
  Class.\ Quant.\ Grav.\  {\bf 36}, no. 20, 205018 (2019)
  [arXiv:1905.09224 [hep-th]].


\bibitem{Puhm:2019zbl}
  A.~Puhm,
  ``Conformally Soft Theorem in Gravity,''
  arXiv:1905.09799 [hep-th].

\bibitem{Guevara:2019ypd}
  A.~Guevara,
  ``Notes on Conformal Soft Theorems and Recursion Relations in Gravity,''
  arXiv:1906.07810 [hep-th].
\bibitem{distler}
  J.~Distler, R.~Flauger and B.~Horn,
  ``Double-soft graviton amplitudes and the extended BMS charge algebra,''
  JHEP {\bf 1908}, 021 (2019)
  [arXiv:1808.09965 [hep-th]].
\bibitem{Osborn:2012vt}
  H.~Osborn,
  ``Conformal Blocks for Arbitrary Spins in Two Dimensions,''
  Phys.\ Lett.\ B {\bf 718}, 169 (2012)
  [arXiv:1205.1941 [hep-th]]

\bibitem{Stieberger:2018onx}
 S.~Stieberger and T.~R.~Taylor,
  ``Symmetries of Celestial Amplitudes,''
  Phys.\ Lett.\ B {\bf 793}, 141 (2019)
  [arXiv:1812.01080 [hep-th]].
\bibitem{Law:2019glh}
  Y.~T.~A.~Law and M.~Zlotnikov,
  ``Poincar\'e Constraints on Celestial Amplitudes,''
  arXiv:1910.04356 [hep-th].
 \bibitem{Osborn1108}
  F.~A.~Dolan and H.~Osborn,
  ``Conformal Partial Waves: Further Mathematical Results,''
  arXiv:1108.6194 [hep-th]
\bibitem{Barnich1703}
  G.~Barnich,
  ``Centrally extended BMS4 Lie algebroid,''
  JHEP {\bf 1706} (2017) 007
  [arXiv:1703.08704 [hep-th]].
\bibitem{Maldacena:2000hw}
  J.~M.~Maldacena and H.~Ooguri,
  ``Strings in AdS(3) and SL(2,R) WZW model 1.: The Spectrum,''
  J.\ Math.\ Phys.\  {\bf 42}, 2929 (2001)
  [hep-th/0001053].
\bibitem{tt}  T.R.~Taylor,
  ``A Course in Amplitudes,''
Phys.\ Rept.\  {\bf 691}, 1 (2017).
[arXiv:1703.05670 [hep-th]].
\bibitem{Stieberger:2015kia}
  S.~Stieberger and T.~R.~Taylor,
  ``Subleading terms in the collinear limit of Yang-Mills amplitudes,''
  Phys.\ Lett.\ B {\bf 750}, 587 (2015)
  [arXiv:1508.01116 [hep-th]].
\bibitem{DeWitt1967}
B.S.\ DeWitt, "Quantum Theory of Gravity. 2. The Manifestly Covariant Theory," Phys. Rev. 162, 11995 ( 1967 )
  \bibitem{Bohr2004}
  N.~E.~Bjerrum-Bohr,
  ``Quantum gravity, effective fields and string theory,''
  hep-th/0410097.
\bibitem{Bern1998}
  Z.~Bern, L.~J.~Dixon, M.~Perelstein and J.~S.~Rozowsky,
  ``Multileg one loop gravity amplitudes from gauge theory,''
  Nucl.\ Phys.\ B {\bf 546}, 423 (1999)
  [hep-th/9811140].
 \end{thebibliography}
\end{document}